\documentclass[fleqn,usenatbib]{mnras}

\usepackage{newtxtext,newtxmath}

\usepackage[T1]{fontenc}

\DeclareRobustCommand{\VAN}[3]{#2}
\let\VANthebibliography\thebibliography
\def\thebibliography{\DeclareRobustCommand{\VAN}[3]{##3}\VANthebibliography}

\usepackage{graphicx}	
\usepackage{amsmath}	
\usepackage{hyperref}

\title[X-ray flux and spectral components of OJ 287 (2005--2020)]{X-ray spectral components of the blazar and binary black hole candidate OJ 287  (2005--2020) \\
}  

\author[S. Komossa et al.]{
S. Komossa,$^{1}$\thanks{E-mail: astrokomossa@gmx.de (SK)}
D. Grupe,$^{2}$
M.L. Parker,$^{3,4}$
J.L. G\'omez,$^{5}$
M.J. Valtonen,$^{6,7}$
M.A. Nowak,$^{8}$
\newauthor
S.G. Jorstad,$^{9,10}$
D. Haggard,$^{11,12}$
S. Chandra,$^{13}$
S. Ciprini,$^{14,15}$
L. Dey,$^{16}$
A. Gopakumar,$^{16}$
\newauthor
K. Hada,$^{17}$
S. Markoff,$^{18}$
J. Neilsen$^{19}$ 
\\
$^{1}$Max-Planck-Institut f\"ur Radioastronomie, Auf dem H{\"u}gel 69, 53121 Bonn, Germany\\
$^{2}$Department of Physics, Earth Science, and Space System Engineering, Morehead State University, 235 Martindale Dr, Morehead, KY 40351, USA\\
$^{3}$European Space Agency (ESA), European Space Astronomy Centre (ESAC), E-28691 Villanueva de la Canada, Madrid, Spain \\
$^{4}$ Institute of Astronomy, University of Cambridge, Madingley Road, Cambridge CB3 0HA, UK \\ 
$^{5}$ Instituto de Astrofísica de Andalucía-CSIC, Glorieta de la Astronomía s/n, E-18008 Granada, Spain, \\
$^{6}$ Finnish Centre for Astronomy with ESO, University of Turku, FI-20014, Turku, Finland \\
$^{7}$ Department of Physics and Astronomy, University of Turku, FI-20014, Turku, Finland \\ 
$^{8}$ Department of Physics, Washington University in St. Louis, One Brookings Dr., St. Louis, MO 63130-4899, USA \\ 
$^{9}$ Institute for Astrophysical Research, Boston University, 725 Commonwealth Ave, Boston, MA 02215, USA \\
$^{10}$ Sobolev Astronomical Institute, St. Petersburg State University, St. Petersburg, Russia \\
$^{11}$ Department of Physics, McGill University, 3600 rue University, Montréal, QC H3A 2T8, Canada \\
$^{12}$ McGill Space Institute, McGill University, 3550 rue University, Montréal, QC H3A 2A7, Canada \\ 
$^{13}$ Centre for Space Research,
North-West University, Potchefstroom 2520, South Africa \\
$^{14}$ Istituto Nazionale di Fisica Nucleare (INFN) Sezione di Roma Tor Vergata, Via della Ricerca Scientifica 1, 00133, Roma, Italy \\
$^{15}$ ASI Space Science Data Center (SSDC), Via del Politecnico, 00133, Roma, Italy \\
$^{16}$ Department of Astronomy and Astrophysics, Tata Institute of Fundamental Research, Mumbai 400005, India \\
$^{17}$ Department of Astronomical Science, The Graduate University for Advanced Studies (SOKENDAI), 2-21-1 Osawa, Mitaka, Tokyo 181-8588, Japan\\
$^{18}$ API–Anton Pannekoek Institute for Astronomy, University of Amsterdam, Science Park 904, 1098 XH Amsterdam, The Netherlands \\
$^{19}$ Villanova University, 800 E. Lancaster Avenue, Villanova, PA 19085, USA \\
 } 

\date{Accepted 2021 April 23. Received 2021 April 16; in original form 2021 January 13}

\pubyear{2020}

\begin{document}
\label{firstpage}
\pagerange{\pageref{firstpage}--\pageref{lastpage}}
\maketitle

\begin{abstract}
We present a comprehensive analysis of all XMM-Newton spectra of OJ 287 spanning 15 years of X-ray spectroscopy of this bright blazar. 
We also report the latest results from our dedicated Swift UVOT and XRT monitoring of OJ 287 which started in 2015, along with all earlier public Swift data since 2005. 
During this time interval, OJ 287 was caught in extreme minima and outburst states. Its X-ray spectrum is highly variable and encompasses all states seen in blazars from very flat to exceptionally steep. 
The spectrum can be 
decomposed into three spectral components: Inverse Compton (IC) emission dominant at low-states, super-soft synchrotron emission which becomes increasingly dominant as OJ 287 brightens, 
and an intermediately-soft ($\Gamma_{\rm x} = 2.2$) additional component seen at outburst. This last component extends beyond 10 keV and plausibly represents either a second synchrotron/IC component and/or a temporary disk corona of the primary supermassive black hole (SMBH).
Our 2018 XMM-Newton observation, quasi-simultaneous with the Event Horizon Telescope observation of OJ 287, is well described by a two-component model 
with a hard IC component of $\Gamma_{\rm x} =1.5$ and a soft synchrotron component.
Low-state spectra limit any long-lived accretion disk/corona contribution in X-rays
to a very low value of $L_{\rm x}/L_{\rm Edd}  < 5.6 \times 10^{-4} $ (for $M_{\rm BH, primary}=1.8 \times 10^{10}$ M$_\odot$).
Some implications for the binary SMBH model of OJ 287 are discussed. 
\end{abstract}



\section{Introduction}

The blazar OJ 287 is perhaps the first identified multi-messenger 
source among extragalactic supermassive black holes. 
As one of the best candidates for hosting a compact supermassive binary black hole (SMBBH) and known as a bright blazar, it has been intensely observed across the {\em{electromagnetic}} regime.
Further, modelling of its optical/IR light curve has revealed evidence for mild orbital shrinkage due to the emission of {\em {gravitational}} waves \citep[GWs;][]{Valtonen2008, Laine2020}.
While not yet directly detected, GWs from this system may become measurable with the next generation of pulsar-timing arrays (PTAs) based on planned radio-telescope facilities
\citep{Yardley2010}.

Coalescing SMBBHs which form in galaxy mergers are the loudest 
sources of low-frequency GWs in
the universe \citep{Centrella2010, Kelley2019}. An intense electromagnetic search for wide and close binaries in all stages of
their evolution is therefore ongoing \citep[review by][]{Komossa2016}. While wide pairs can be identified by spatially-resolved imaging
spectroscopy, indirect methods are required for detecting the most compact, evolved systems. 
These latter systems are well beyond the ``final parsec'' in their evolution 
\citep{Begelman1980, Colpi2014} and  
in a regime where GW emission contributes to, or dominates, their orbital shrinkage.
Semi-periodicity in light curves has become a major
tool for selecting these small-separation SMBBH candidates \citep[e.g.,][]{Graham2015}. 

OJ 287 is a nearby, bright BL Lac object at redshift $z=0.306$ 
and among the best candidates to date
for hosting a compact SMBBH \citep{Sillanpaa1988, Valtonen2016}. 
Its unique optical light curve spans more than a century and dates back to the late 19th century.  
It shows optical double-peaks every $\sim$12 years, which
have been interpreted as arising 
from the orbital motion of a pair of SMBHs, with
an orbital period on that order ($\sim$9 yrs in the system's rest frame). 

While different variants of binary scenarios of OJ 287 were considered 
in the past \citep[e.g.,][]{Lehto1996,  Katz1997, Valtaoja2000, Villata1998, 
Liu2002, Qian2015, Britzen2018, Dey2018}, 
the best explored and most successful model explains the double peaks as 
the times when the secondary SMBH impacts the disk around the primary twice during its $\sim$12 yr
orbit ("impact flares" hereafter). The most recent orbital two-body modelling 
is based on 4.5 order post-Newtonian dynamics and successfully reproduces the
overall long-term light curve of OJ 287 until 2019 \citep[][and references therein]{Valtonen2016, Dey2018, Laine2020}, with impact flares observed most recently in 2015 and 2019. The model requires a compact binary with a semi-major axis of 9300 AU 
with a massive primary SMBH of $1.8\times10^{10}$
M$_{\odot}$, and a secondary of $1.5\times10^8$ M$_{\odot}$. 
Because of the strong general-relativistic precession of the secondary's orbit, $\Delta \Phi$=38 deg/orbit, the impact flares are not always separated by 12 yrs, instead their separation varies strongly with time in a predictable manner. 
In addition to the impact flares, the model predicts "after-flares" when the impact disturbance reaches the inner accretion disk \citep{Sundelius1997, Valtonen2009, Pihajoki2013}, identified most recently with the bright X-ray--UV--optical outburst in 2020 \citep{Komossa2020a}.

First detected as a radio source during the Vermilion River Observatory Sky Survey \citep{Dickel1967}
and the Ohio Sky Survey \citep{Kraus1977},
OJ 287 has been studied extensively in the radio regime.
Its relativistic jet is pointing at us with an average viewing angle of $\sim2$ degrees \citep{Jorstad2005, Agudo2012} and shows remarkable short-timescale variability interpreted as a turbulent injection process and/or a clumpy accretion disk structure 
\citep{Agudo2012} or as a binary-induced wobble \citep{ValtonenPihajoki2013, Dey2021}.
The inner jet is the source of highly variable $\gamma$-rays \citep[e.g.,][]{Hodgson2017, Agudo2011} and displays strong and variable radio polarization  \citep[e.g.,][]{Aller2014, Myserlis2018, Cohen2018}. 
Though only faintly detected in the very-high energy (VHE) band \citep[$>$100 GeV;][] {Mukherjee2017, O'Brien2017}, 
OJ 287 is a well-known X-ray emitter and has been detected with most major X-ray observatories, including 
Einstein \citep{Madejski1988}, 
EXOSAT \citep{Sambruna1994}, 
ROSAT \citep{Comastri1995},
BeppoSAX \citep{Massaro2003}, 
ASCA \citep{Idesawa1997},
the Neil Gehrels Swift observatory
\citep[Swift hereafter;][]{Massaro2008}, 
Suzaku \citep{Seta2009}, 
XMM-Newton \citep{Ciprini2007} and most recently with NuSTAR \citep{Komossa2020a}.
These observations established OJ 287 as a bright and variable X-ray source, and allowed single-component X-ray spectral fits.
Its (0.5--10) kev X-ray spectrum was interpreted as a mix of synchrotron and inverse Compton (IC) emission, with the former more variable \citep{Urry1996}. OJ 287 is found most of the time in a rather flat spectral state with a photon index $\Gamma_{\rm x} \approx 1.5-1.9$, though $\Gamma_{\rm x}$ is as steep as 2.6 during one ROSAT observation \citep{Urry1996} in the band (0.1--2.4) keV.   
Its steepest state, with an equivalent $\Gamma_{\rm x} = 2.8$, 
(better fit with a logarithmic parabolic model)
was detected with XMM-Newton in the band (0.3--10) keV, which caught the source right at the peak of one of the brightest X-ray outbursts measured so far \citep{Komossa2020a}. 
Imaging with the Chandra X-ray observatory has revealed a long, curved X-ray jet consisting of multiple knots and extending out to 20\arcsec or a de-projected scale of $>1$ Mpc, and bright central emission \citep{Marscher2011}. 

The optical spectrum of OJ 287 is often featureless, even though some narrow emission lines and broad H$\alpha$ are occasionally detected in low-states and have been used to determine the redshift of OJ 287 \citep{SitkoJunkkarinen1985, Stickel1989}. H$\alpha$ had decreased by a factor of 10 when re-observed in 2005-2008 \citep{Nilsson2010}. 
The overall faintness of the broad Balmer lines is the basis for the classification of OJ 287 as a BL Lac object. 

We are carrying out a dedicated multi-year, multi-frequency monitoring program of OJ 287
with Swift and in the radio regime.
This MOMO program (for Multi-wavelength Observations and Modelling of OJ 287) includes a search for outbursts and explores and tests facets of the binary SMBH model. Previous results from this program were presented by \citet[][flux and spectral evolution during the 2016/17 outburst]{Komossa2017}, by \citet[][radio monitoring]{Myserlis2018}, and by  \citet[][2020 outburst and long-term lightcurve; "paper I" hereafter]{Komossa2020a}.  Exceptional outbursts or low-states of OJ 287 detected by Swift were initially reported within days in a sequence of {\sl Astronomer's Telegrams} \citep[e.g.,][]{Komossa2015, Grupe2016, Grupe2017, Komossa2018, Komossa2020b, Komossa2020c, Komossa2020d} to inform the community and encourage observations in bands not densely covered by our campaigns, like the near infrared \citep{Kushwaha2018}.  
Independent of the binary's presence, OJ 287 is a nearby bright blazar, and dense multi-frequency monitoring and high-resolution X-ray spectroscopy are powerful diagnostics of jet and accretion physics in blazars.
Results of the MOMO program continue to be presented in a sequence of publications. Here, in the fourth study, we present  XMM-Newton 
and Swift X-ray spectroscopy of OJ 287, spanning a time interval of 1.5 decades and covering the most extreme X-ray flux and spectral states so far observed. 

\begin{figure*}
\includegraphics[clip, trim=1.8cm 5.6cm 1.3cm 2.6cm, width=17.8cm]{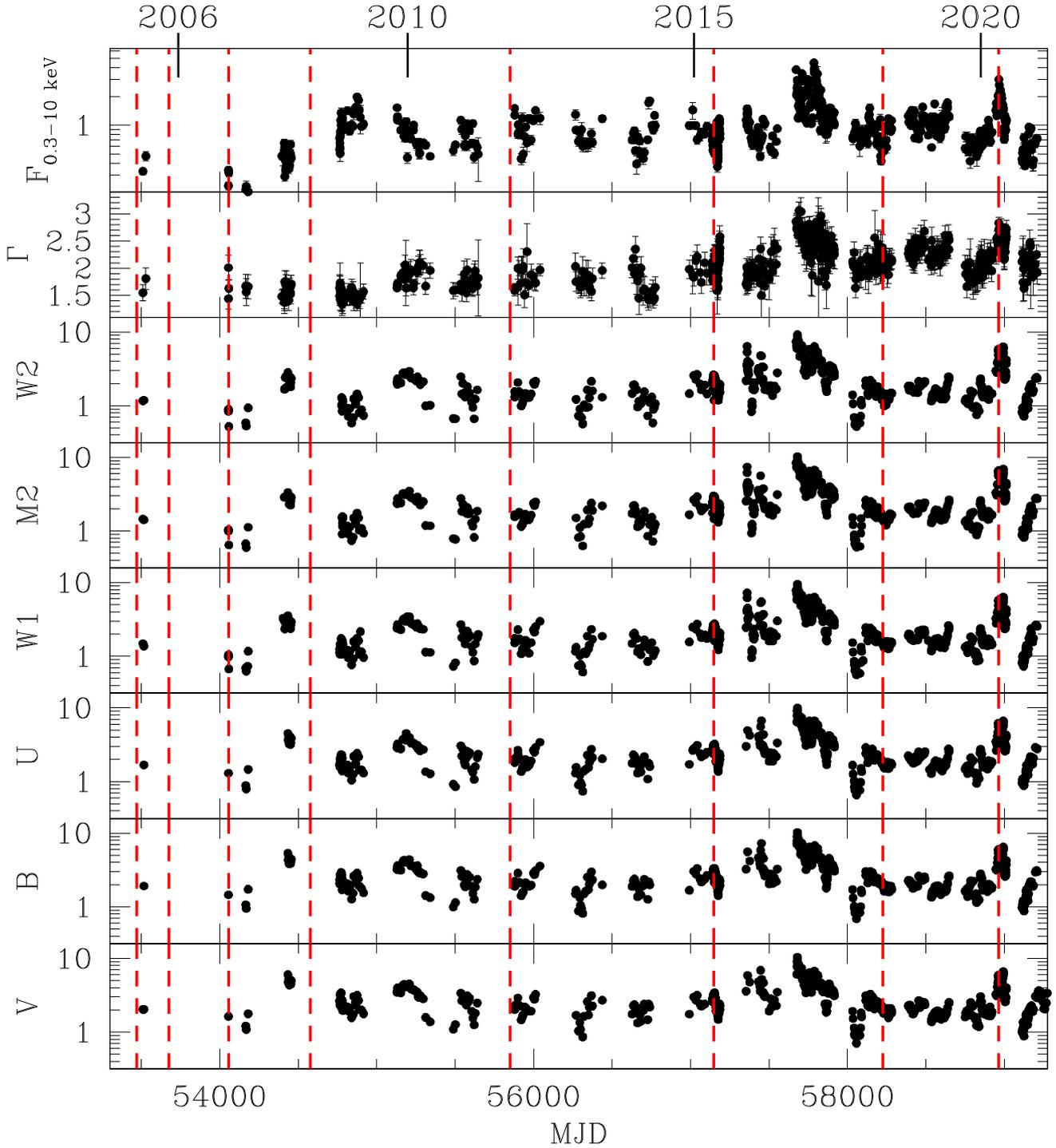}
    \caption{Swift XRT and UVOT light curves of OJ 287 between 2005 and December 2020 since the beginning of Swift monitoring. 
    Epochs of XMM-Newton observations are marked by vertical, dashed, red lines.
    The observed X-ray flux (0.3--10 keV, corrected for Galactic absorption) and the optical-UV fluxes (corrected for extinction) are given in units of  10$^{-11}$ erg\,s$^{-1}$\,cm$^{-2}$. $\Gamma_{\rm x}$ is the X-ray power-law photon index. 
    January 1st of each of the years 2006, 2010, 2015 and 2020 is marked on the upper horizontal axis. 
Error bars are always reported, but are often smaller than the symbol size.
}
\label{fig:lc-Swift}
\end{figure*}

This paper is structured as follows. In Sect. 2 we discuss the long-term Swift data with focus on the X-ray spectral variability properties and a softer-when-brighter variability pattern. A spectral and temporal analysis of all XMM-Newton data of OJ 287 is provided in Sect. 3. In Sect. 4, implications for emission models and the binary SMBH nature of OJ 287 are discussed. A summary and conclusions are provided in Sect. 5.
A cosmology \citep{Wright2006} with 
$H_{\rm 0}$=70 km\,s$^{-1}$\,Mpc$^{-1}$, $\Omega_{\rm M}$=0.3 and $\Omega_{\rm \Lambda}$=0.7 is used throughout this paper. 

\section{Swift data analysis and spectral fits}

%
\begin{table}
\footnotesize
	\centering
	\caption{Log of observations of our new 2020 Swift data since September 2020. The third column gives the dates of the observations and the fourth column reports the duration of each single observation in ks.
	}
	\label{tab:obs-log}
	\begin{tabular}{lclc}
		\hline
		instrument & band/filter & date & $\Delta t$ (ks)\\
		\hline
		XRT & 0.3--10 keV & Sept.--Dec. 2020 & 0.5--2 \\
        UVOT & W2 & Sept.--Dec. 2020 & 0.12--0.6 \\
             & M2 & & 0.09--0.5 \\
             & W1 & & 0.06--0.3 \\
             & U & & 0.03--0.16 \\
             & B & & 0.03--0.16 \\
             & V & & 0.03--0.16 \\
		\hline
	\end{tabular}
\end{table}

\begin{figure*}
\includegraphics[clip, trim=2.0cm 5.6cm 1.3cm 3.1cm, width=16.0cm]{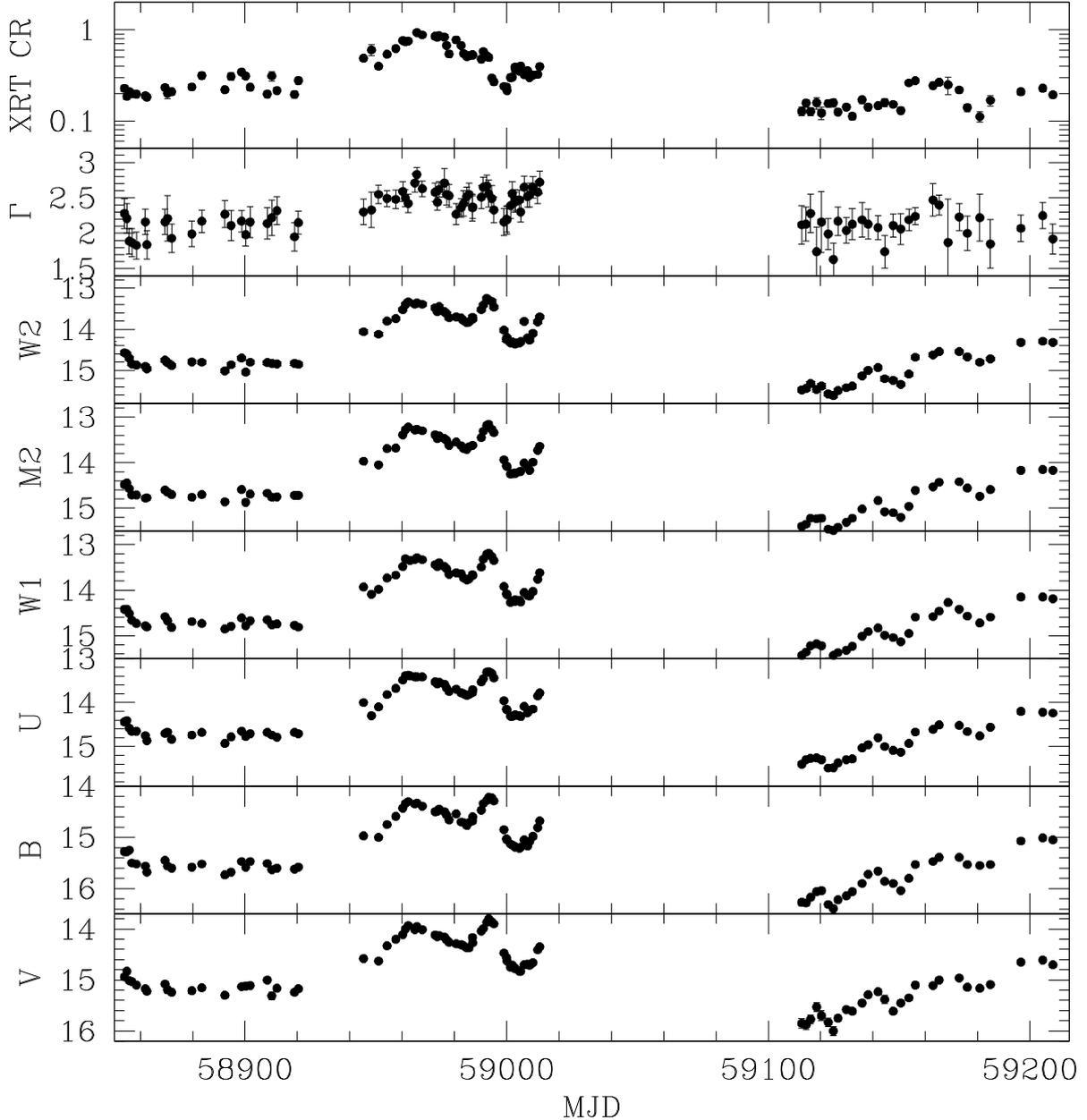}
    \caption{2020 Swift light curve (XRT countrate in cts\,s$^{-1}$, photon index $\Gamma_{\rm x}$, and observed UVOT magnitudes in the VEGA system) of OJ 287 obtained in the course of the MOMO program, 
    including the April--June 2020 outburst \citep{Komossa2020a} and the newly detected deep September 2020 optical-UV low-state when OJ 287 emerged from Swift Sun constraint. Error bars are always reported, but are often smaller than the symbol size.
    }
    \label{fig:zoom-Swift}
\end{figure*}

\subsection{Swift XRT} 
We have monitored OJ 287 with Swift \citep{Gehrels2004}
since December 2015 \citep[][our Fig. \ref{fig:lc-Swift}]{Komossa2017, Komossa2020a}. A cadence of $\sim$3--7 days was employed and increased to 1--2 days at some epochs of outbursts or low-states. We also enhanced the already dense monitoring cadence of OJ 287 during the epochs of the Event Horizon Telescope \citep[EHT hereafter;][]{EHT2019}  coverage in April 2017 and 2018.

Our monitoring was resumed in September 2020 
(Tab. \ref{tab:obs-log}, 
Fig. \ref{fig:zoom-Swift}; OBS-ID 35905-89...35905-117), when OJ 287 became visible again after its Swift Sun constraint. We found that the bright April--May 2020 outburst was over, and OJ 287 was caught in a deep low-state at all wavebands. The light curve is reported here for the first time. The community was informed about the low-state in an {\sl{Astronomer's Telegram}} \citep{Komossa2020d}. 
During these new 2020 observations, the Swift X-ray telescope \citep[XRT;][]{Burrows2005} was operating in {\sl{photon counting}} (PC) mode
\citep{Hill2004}. 
Exposure times range between 
0.5--2 ks. 

To inspect the long-term light curve, and obtain the X-ray and UVOT flux evolution before and after the epochs of the early XMM-Newton observations (Sect. 3) as well, we have also created the Swift UVOT and XRT light curve of OJ 287 between 2005 and 2015 based on all public archival data. 
Further, we have added all of the few data after 2015 which were not part of our own dedicated monitoring program. 
Before late 2015 the Swift coverage of OJ 287 was overall more sparse, 
with the exception of a dense monitoring campaign in mid-2015 (PI: R. Edelson). 
Many of the remaining pre-2015 data were obtained in the course of a blazar monitoring program \citep{Massaro2008, StrohFalcone2013, Williamson2014, Siejkowski2017} and a dedicated early study of OJ 287 \citep{Valtonen2016}. The majority of these XRT data were
taken in PC mode, while data above $\sim$1 cts s$^{-1}$ were obtained in {\sl{windowed timing}} (WT) mode \citep{Hill2004}.

X-ray countrates were determined using the XRT product tool at the Swift data centre in Leicester \citep{Evans2007}. 
To carry out the X-ray spectral analysis, source photons were extracted within a circle of 47\arcsec~radius.
OJ 287 is off-axis in most observations, as is typical for Swift monitoring programs.  The source extraction size does include the 20\arcsec~X-ray jet detected with the Chandra observatory.  However, the integrated Chandra ACIS-S jet emission of $\sim$0.03 cts s$^{-1}$ with an average $\Gamma_{\rm x} = 1.61$ \citep{Marscher2011} only amounts to a Swift XRT countrate of 0.009 cts s$^{-1}$, negligibly contributing to the integrated emission even in X-ray low-states.
Background photons were collected in a nearby circular region of radius 236\arcsec. 
X-ray spectra in the band (0.3--10) keV were then generated 
and the software package \textsc{xspec} \citep[version 12.10.1f;][]{Arnaud1996} was used for analysis. 

Swift data above a countrate of $\sim$0.7 cts s$^{-1}$ 
are affected by pile-up. To correct for it,  the standard procedure 
was followed of first creating a region file where the inner circular area of the PSF 
was excluded from the analysis. The loss in counts is then corrected by creating a new ancillary response file based on this annular region that is used in \textsc{xspec} to correct the flux measurement.

Spectra were fit with single power laws of photon index $\Gamma_{\rm X}$ 
adding Galactic absorption 
\citep{Wilms2000} along our line-of-sight 
with a Hydrogen column density $N_{\rm H, Gal}=2.49\times10^{20}$ cm$^{-2}$ \citep{Kalberla2005}, and using 
the W-statistics of \textsc{xspec} 
on the unbinned data.  
(W-statistics are used 
when countrates are low so that data follow Poisson statistics, and when not only a source but also a background spectrum is available.)

\begin{figure}
\includegraphics[clip, angle=270, trim=0.7cm 2.2cm 1.7cm 0.5cm,  width=\columnwidth]{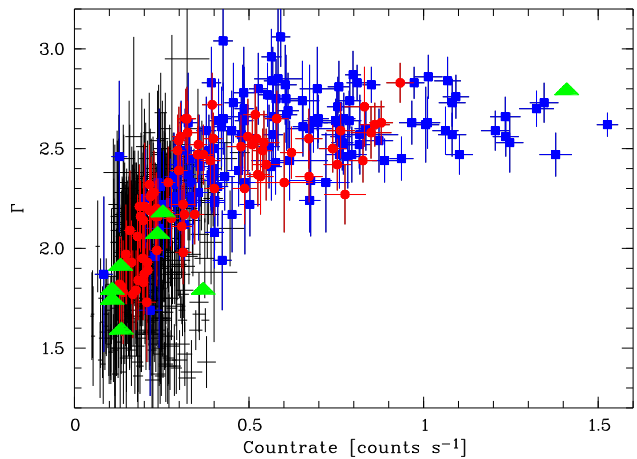}
    \caption{Variability of the photon index $\Gamma_{\rm x}$ of OJ 287 vs. Swift XRT countrate. A softer-when-brighter pattern is apparent (Sect. 2.1), confirming results of paper I.
    The time intervals including the two super-outbursts of OJ 287, in 2016-2017 and in 2020, are overplotted in blue 
    (MJD 57673 -- 58000) 
    and red 
    (MJD 58800 -- 59100), 
    respectively.  Observations between 2005 and end of November 2020 are included in this plot. The eight XMM-Newton data sets were added for comparison (green triangles).
    }
    \label{fig:Swift-HR}
\end{figure}

OJ 287 displays a strong softer-when-brighter variability pattern, shown in Fig. \ref{fig:Swift-HR}. During outbursts, the photon index increases as the countrate rises, and then saturates at the highest countrates. OJ 287 is never found in a flat-spectrum state when brightest. 
A correlation analysis for the whole data set gives a Spearman rank order correlation 
coefficient $r_{\rm s} = 0.61$ and a Student's T-test of $T_{\rm S} = 23$. 
For $N$=681 observations, this corresponds to a probability of a random result of $P<10^{-8}$. 
For the epoch including the 2016/17 outburst (blue data points of Fig. \ref{fig:Swift-HR}) we find $r_{\rm s} = 0.55 , T_{\rm S} = 7.4$,
and $P<10^{-8}$ ($N$=125),
and for the 2020 outburst (red data points) we obtain $r_{\rm s} = 0.8, T_{\rm S} = 11.8$,  
and $P<10^{-8}$ ($N$=79).
This analysis shows that the data are correlated. Always, the probability of a random result is $P<10^{-8}$.

\subsection{Swift UVOT}

We also observed OJ 287 with the UV-optical telescope \citep[UVOT;][]{Roming2005}  in all six filters 
[V (5468\AA), B(4392\AA), U(3465\AA),  UVW1(2600\AA), UVM2(2246\AA), UVW2(1928\AA); where values in brackets are the filter central wavelengths] since the end of 2015 to obtain spectral energy distribution information of this rapidly varying blazar. Public archival data since 2005 were added to the analysis.

In each UVOT filter the observations were co-added.
Source counts in all six filters were then selected in a circle of radius 
5\arcsec ~while the background was determined in a nearby region of radius 20\arcsec.
The background-corrected counts were converted into 
fluxes based on the latest calibration as described in \citet{Poole2008} and \citet{Breeveld2010}. 
In particular, the latest CALDB update version 20200925 which affects data since 2017 was employed.{\footnote{\url{https://www.swift.ac.uk/analysis/uvot/index.php}}}
The UVOT data were corrected for Galactic reddening with $E_{\rm{(B-V)}}$=0.0248 \citep{Schlegel1998}, using a correction factor in each filter according to Equ. (2) of \citet{Roming2009} and based on the reddening curves of \citet{Cardelli1989}.

\begin{table*}
	\centering
	\caption{Summary of XMM-Newton X-ray observations of OJ 287, where OBS-ID is the observation identification number, $\Delta t_{\rm total}$ is the observation duration in ks, and $\Delta t_{\rm eff}$ is the effective exposure time after epochs of flaring particle background were removed. 
	}
	\label{tab:obs-log-XMM}
	\begin{tabular}{lcccccl}
		\hline
		observation mode & OBS-ID & date & MJD & $\Delta t_{\rm total}$ (ks) & $\Delta t_{\rm eff}$ (ks) & comment \\
		\hline
		large-window & 0300480201 &  2005-04-12 & 53472 & 18.5 & 4.6 & hardest state, low flux level \\
		large-window & 0300480301 & 2005-11-03 & 53677 & 39.0 & 23 &  \\
		large-window & 0401060201 & 2006-11-17 & 54056 & 45.0 & 42 &  \\
		large-window & 0502630201 & 2008-04-22 & 54578 & 53.6 & 36 &  \\
		large-window & 0679380701 &  2011-10-15 & 55849 &  21.7 & 20 & brightest state at $E>5$ keV\\
		large-window & 0761500201 & 2015-05-07 & 57149 & 121.4 & 60 & \\
		large-window & 0830190501 & 2018-04-18 & 58226 & 22.3 & 10.5 & quasi-simultaneous with EHT observation \\
small-window & 0854591201 & 2020-04-24 & 58963 & 13.3 & 9 & super-soft state; highest flux level, taken at peak of 2020 outburst \\ 
		\hline
	\end{tabular}
\end{table*}

\section{XMM-Newton data analysis and spectral fits}

\subsection{Data preparation}

We have analyzed all XMM-Newton \citep{Jansen2001} observations of OJ 287 (Tab. \ref{tab:obs-log-XMM}). This includes our PI data from 2018--2020 (PI: S. Komossa/N. Schartel) and from 2005--2011 (PI: S. Ciprini) and an observation from 2015 (PI: R. Edelson). 

While some XMM-Newton observations have been published before \citep[e.g.,][]{Ciprini2007, Valtonen2012, Gaur2018, Gallant2018, Pal2020, Komossa2020a}, here, we carry out a systematic analysis and comparison of all data sets in the context of the same model prescriptions and using the latest calibration data, and with some focus on our 2018 observation of April 18. That observation was carried out quasi-simultaneous with the EHT observation of OJ 287 on 2018 April 26-27 and with the GMVA+ALMA observation on 2018 April 15. 

\begin{figure*}
\includegraphics[width=10.5cm]{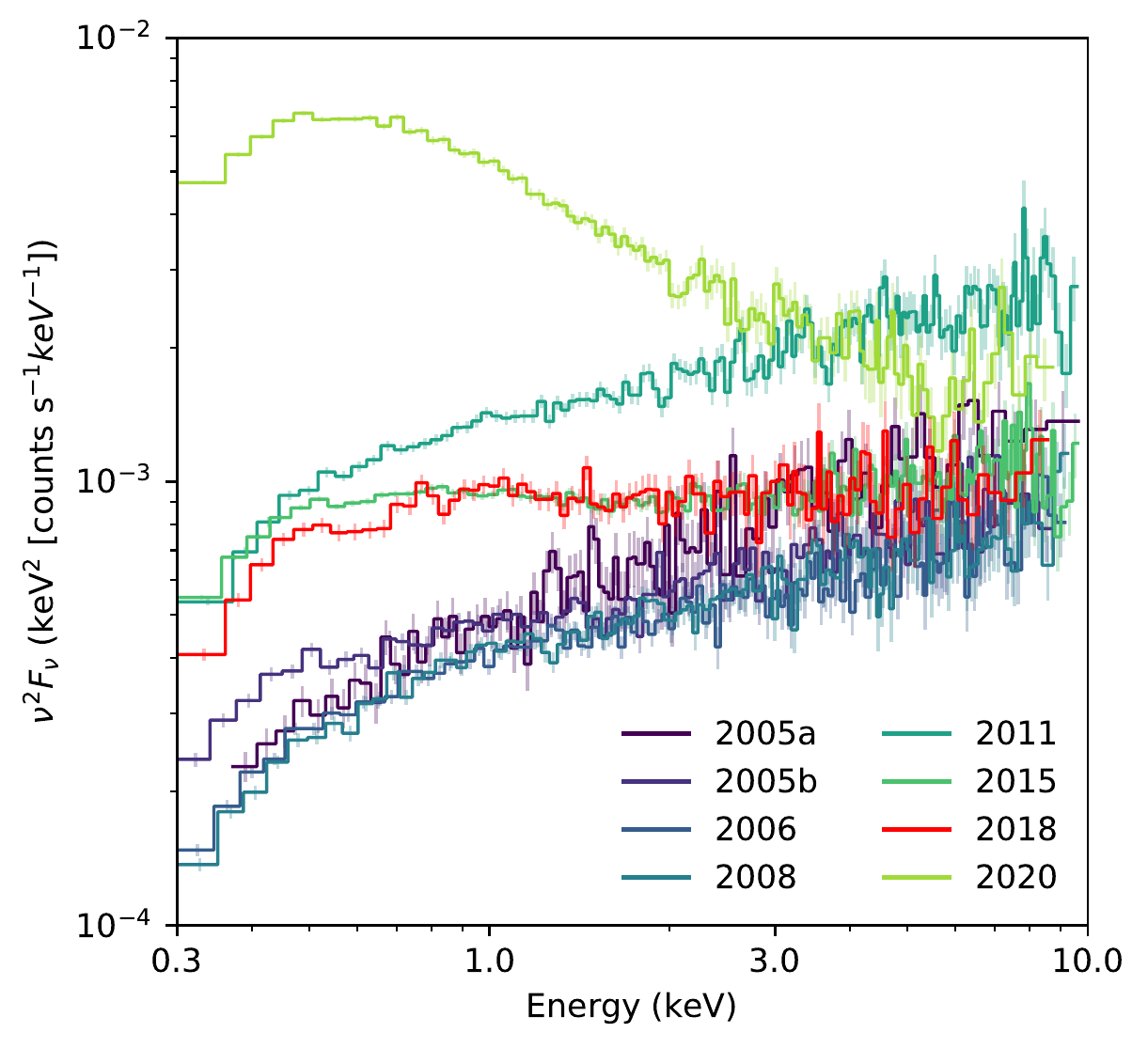}
    \caption{
     Comparison of all XMM-Newton EPIC-pn spectra of OJ 287 in the observer's frame between 2005 and 2020, corrected for the effective area of the detector and uncorrected for Galactic absorption.
     Our 2018 data are highlighted in red, while other observations are marked with different colors and labelled by year of observation. Two spectral states are apparent. A "hard state" dominated by flat power-law emission, and a "soft state" with an additional low-energy component which becomes increasingly dominant when OJ 287 is X-ray bright. The super-soft emission component strongly dominates the 2020 outburst spectrum.
     }
    \label{fig:XMM-allobs}
\end{figure*}

The observations span a wide range in spectral and flux states (Fig. \ref{fig:XMM-allobs}). The majority of them was carried out in large-window mode  
when OJ 287 was at low-intermediate brightness states. The 2020 high-state observation was acquired in small-window mode. In these partial-window modes, only a part of the whole CCD is used to collect data in order to minimize the effect of photon pile-up. 

The XMM-Newton data were reduced using the Science Analysis Software (SAS) version 18.0.0. 
Data were checked for epochs of flaring particle background, and the corresponding time intervals were excluded from the analysis resulting in effective exposure times listed in Tab. \ref{tab:obs-log-XMM}.  
EPIC-pn and EPIC-MOS spectra \citep{Strueder2001} were
extracted in a circular region of 
20\arcsec~radius centered on the 
source position, while background photons were collected in a nearby
region of $\sim50$\arcsec~for the pn and $\sim100$\arcsec~for the MOS instruments. 
Reflection Grating Spectrometer \citep[RGS;][]{denHerder2001} data of the 2018 observation were inspected. We used \textsc{rgsproc} with standard settings to extract the spectra and background for each detector, and combined the first order spectra from the two instruments into a single higher-signal spectrum. The pn and MOS data were binned to a signal-to-noise ratio of at least 6, and to oversample the instrumental resolution by a factor of at least 3.
The RGS spectrum was rebinned by a factor of 8.

\subsection{Spectral fits}

To carry out the spectral analysis, spectra were  
fit with several emission models (Tab. ~\ref{tab:spec-fits}) which are typical for the X-ray spectra of blazars, including a power law, a power law plus soft excess emission parametrized either as a black body component or second power law or a curved logarithmic parabolic power law model (logpar hereafter). 
These latter models are power laws with an index which varies as a log parabola in energy \citep{Massaro2004}, 
with $N(E) = N_{\rm 0} \, (E/E_1)^{-(\alpha + \beta \, log(E/E_1))}$,
where $N$ is the number of photons, $N_{\rm 0}$ is the normalization, $E_1$ is the pivot energy fixed at 1 keV, $\beta$ is the curvature parameter, and $\alpha$ is the slope at the pivot energy. 
The various models are chosen with different scenarios in mind: the single power law is the most common emission model and visualizes the deviations from such a simple model, if any. In two-component models, a power law is always included to represent the high-energy emission component of OJ 287 typically of inverse Compton origin.
A black body emission component is added to represent any kind of soft excess, including possibly (but unlikely) from any temporary accretion disk of the secondary SMBH. The logpar model is representative of a sum of different synchrotron components and is commonly used in the analysis of blazar spectra.   
Neutral absorption along our line-of-sight within our Galaxy is always included \citep[modeled with 
TBnew{\footnote{\url{https://pulsar.sternwarte.uni-erlangen.de/wilms/research/tbabs/}}};][]{Wilms2000}. In single-component power-law fits, additional absorption at $z=0.306$ was added and left free to vary. {\footnote{The most common source of absorption in blazars is the interstellar medium of the host galaxy. In rare cases, single BLR (or NLR) clouds of large column density can be located along the line of sight between the jet(knot) and the observer. Other sources of absorption (ionized or neutral) can be accretion-disk winds or ionized outflows driven by jet-cloud interactions.}}
$\chi^2$ statistics were used to evaluate the goodness of the fit when carrying out the X-ray spectral analysis.

\subsection{Spectral fit results} 

\begin{figure*}
\includegraphics[width=0.32\linewidth]{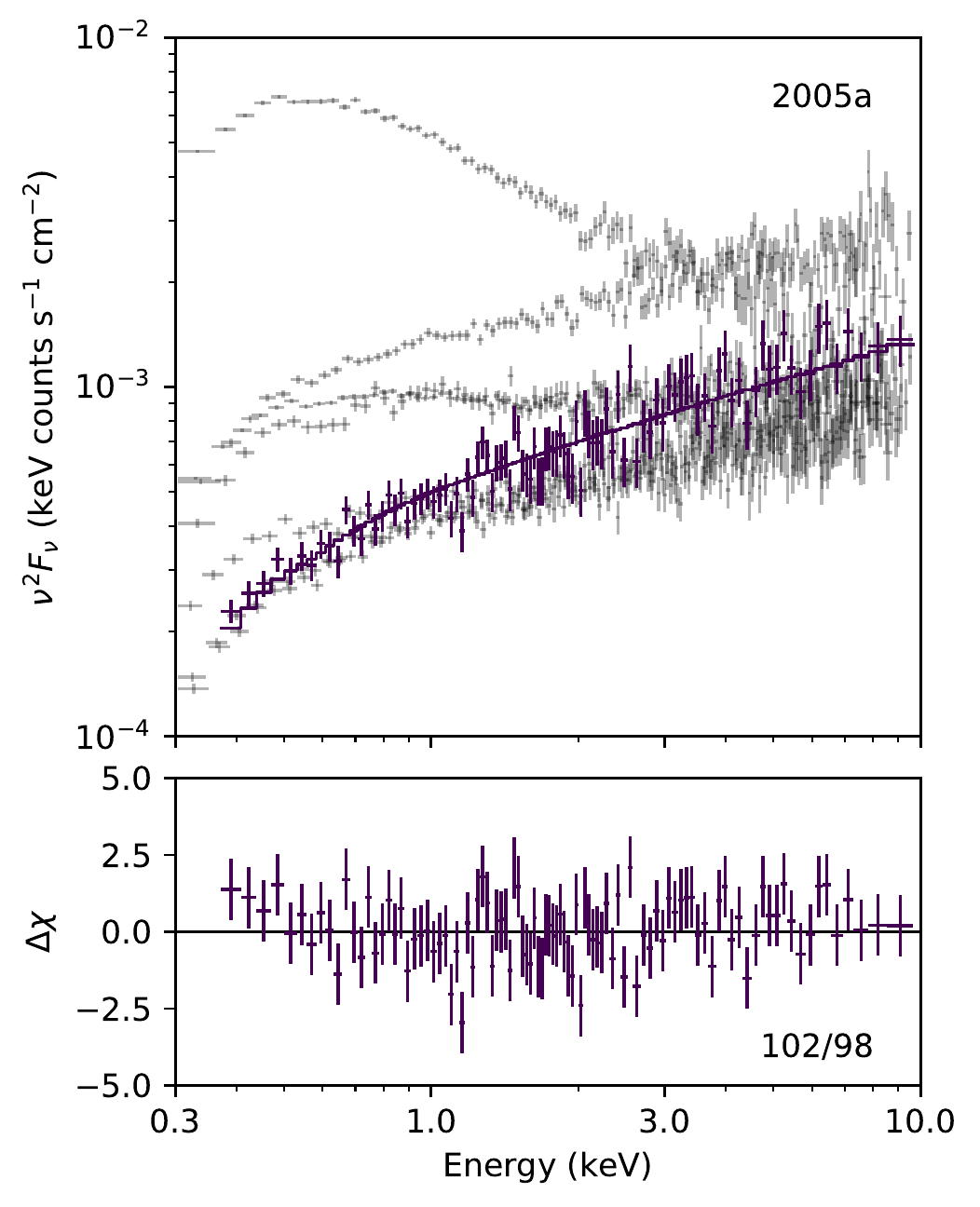}
\includegraphics[width=0.32\linewidth]{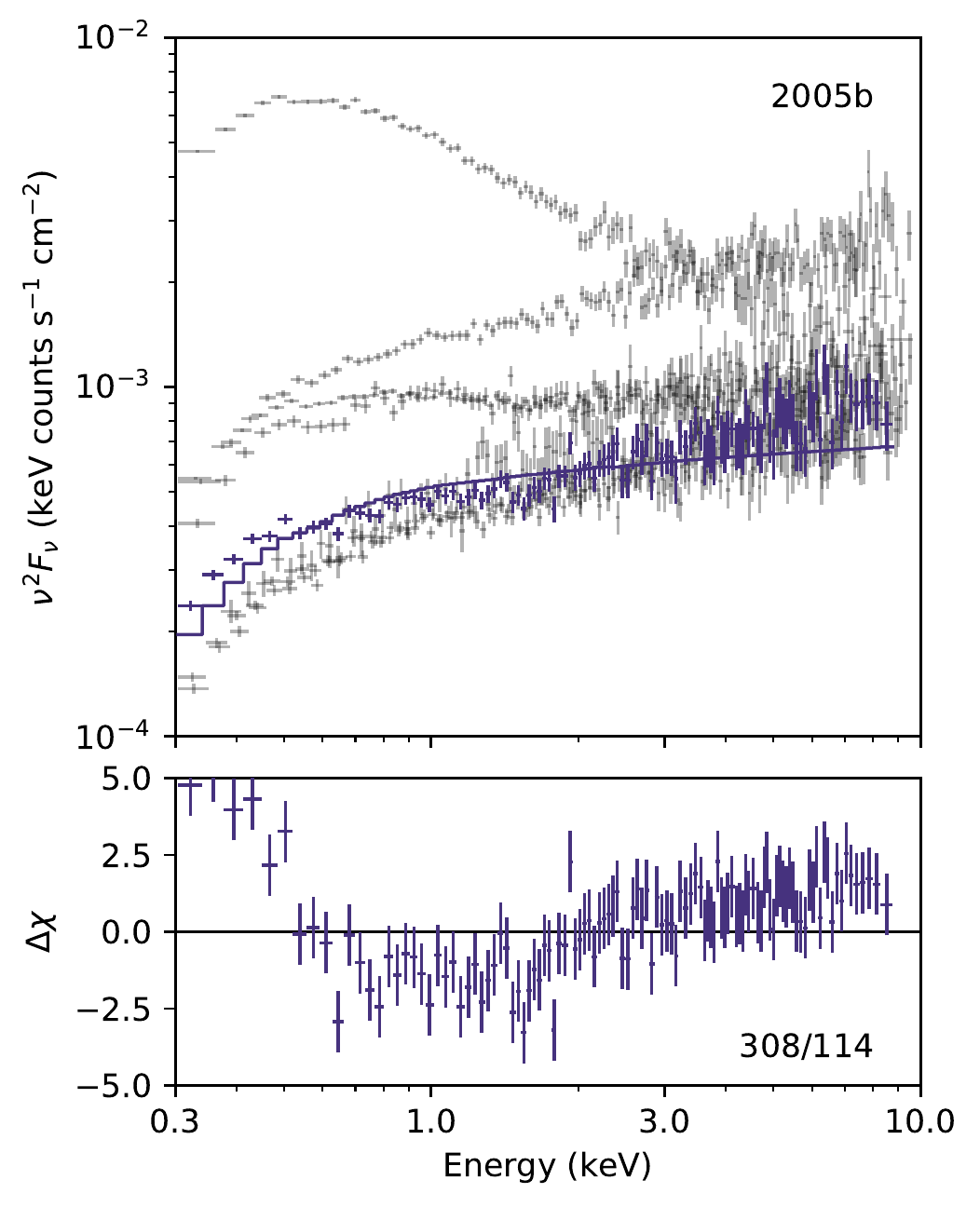}
\includegraphics[width=0.32\linewidth]{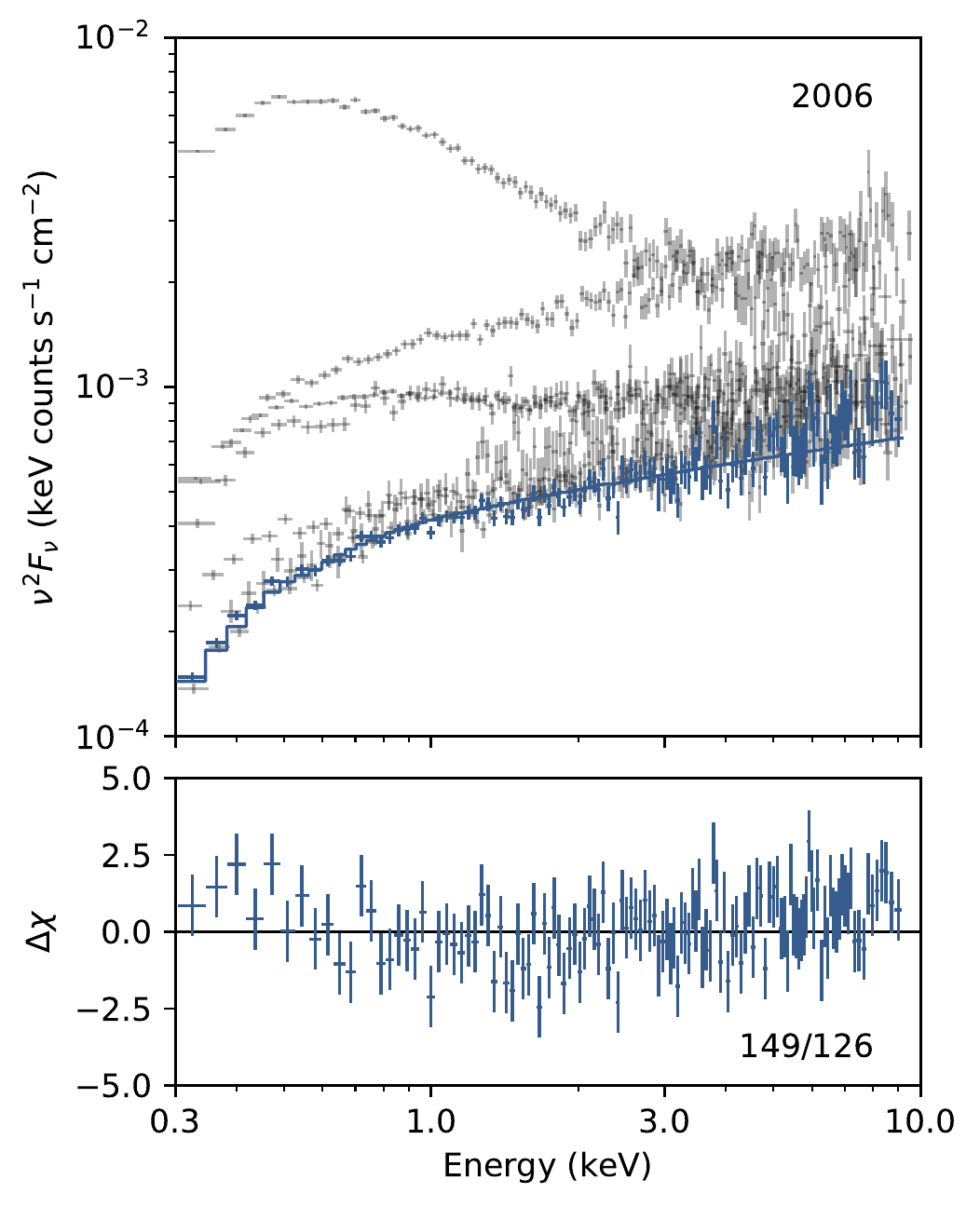}
\includegraphics[width=0.32\linewidth]{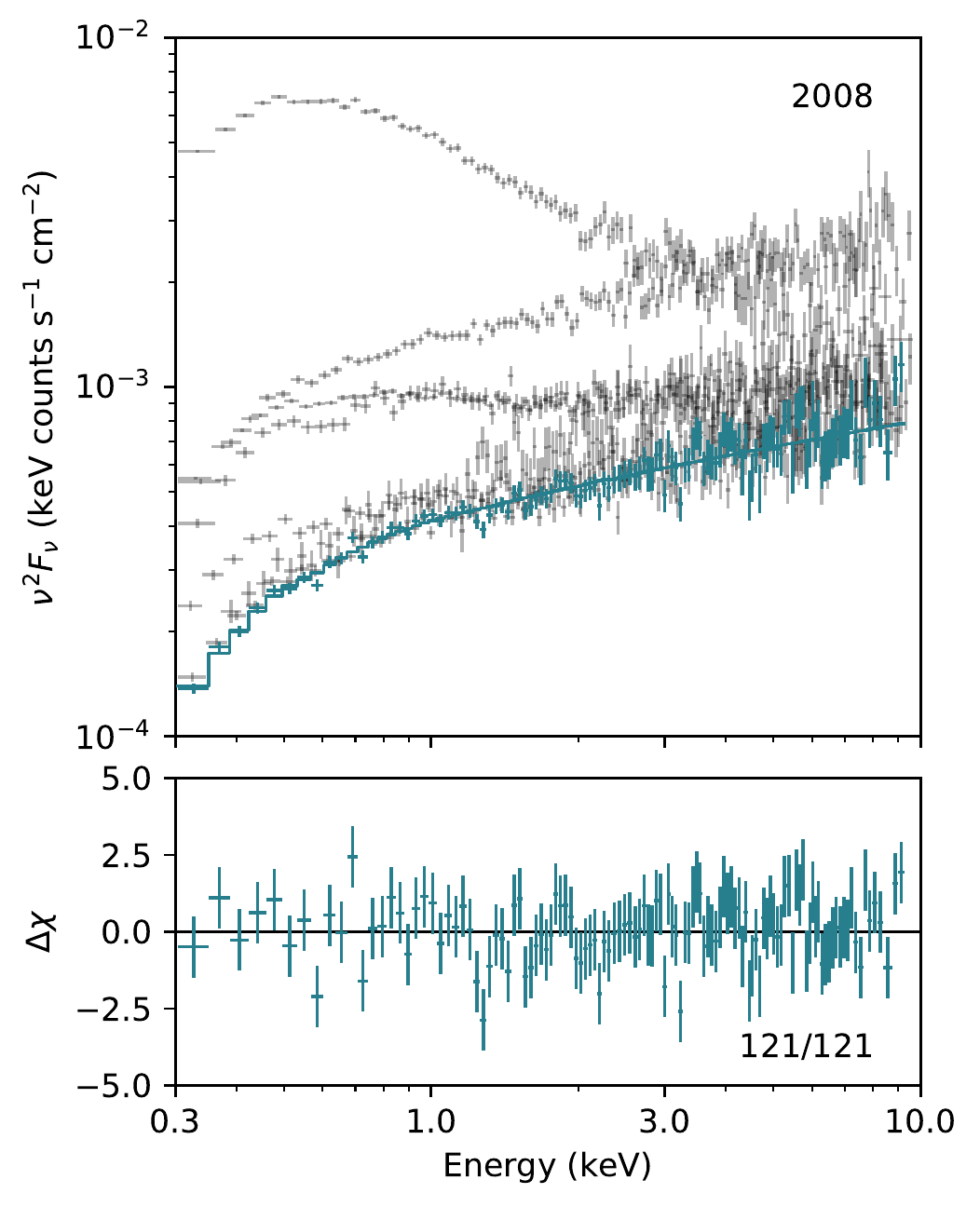}
\includegraphics[width=0.32\linewidth]{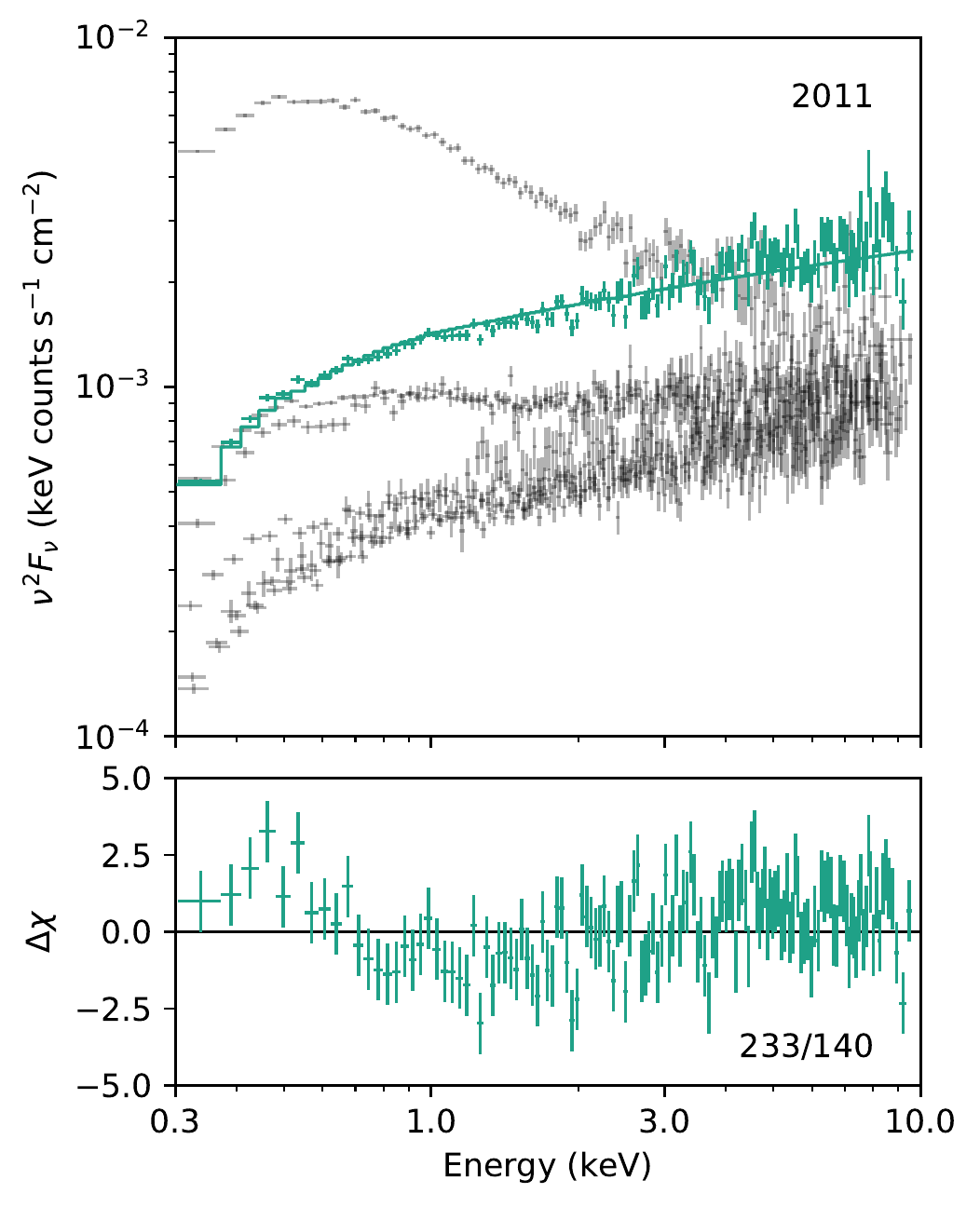}
\includegraphics[width=0.32\linewidth]{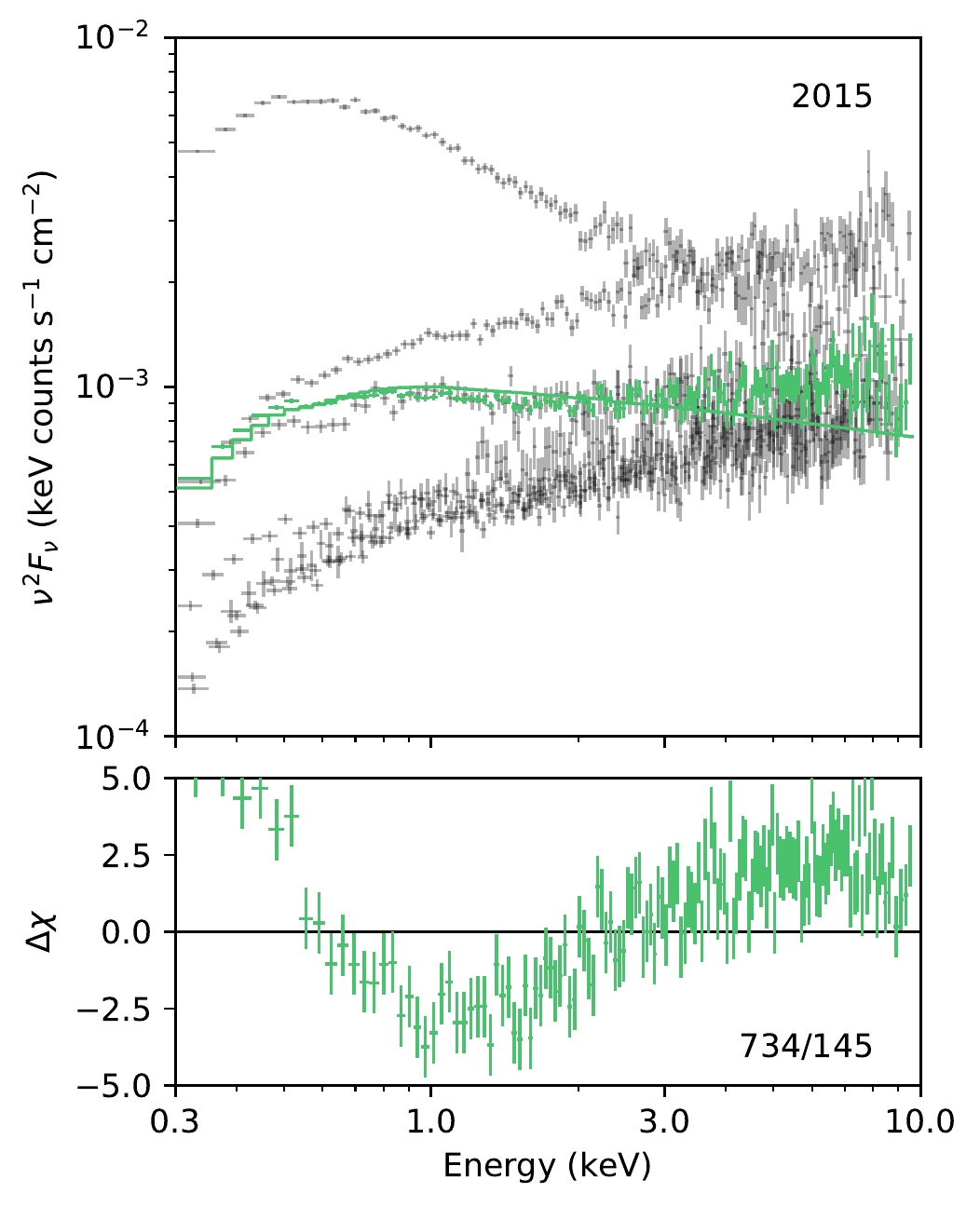}
\includegraphics[width=0.32\linewidth]{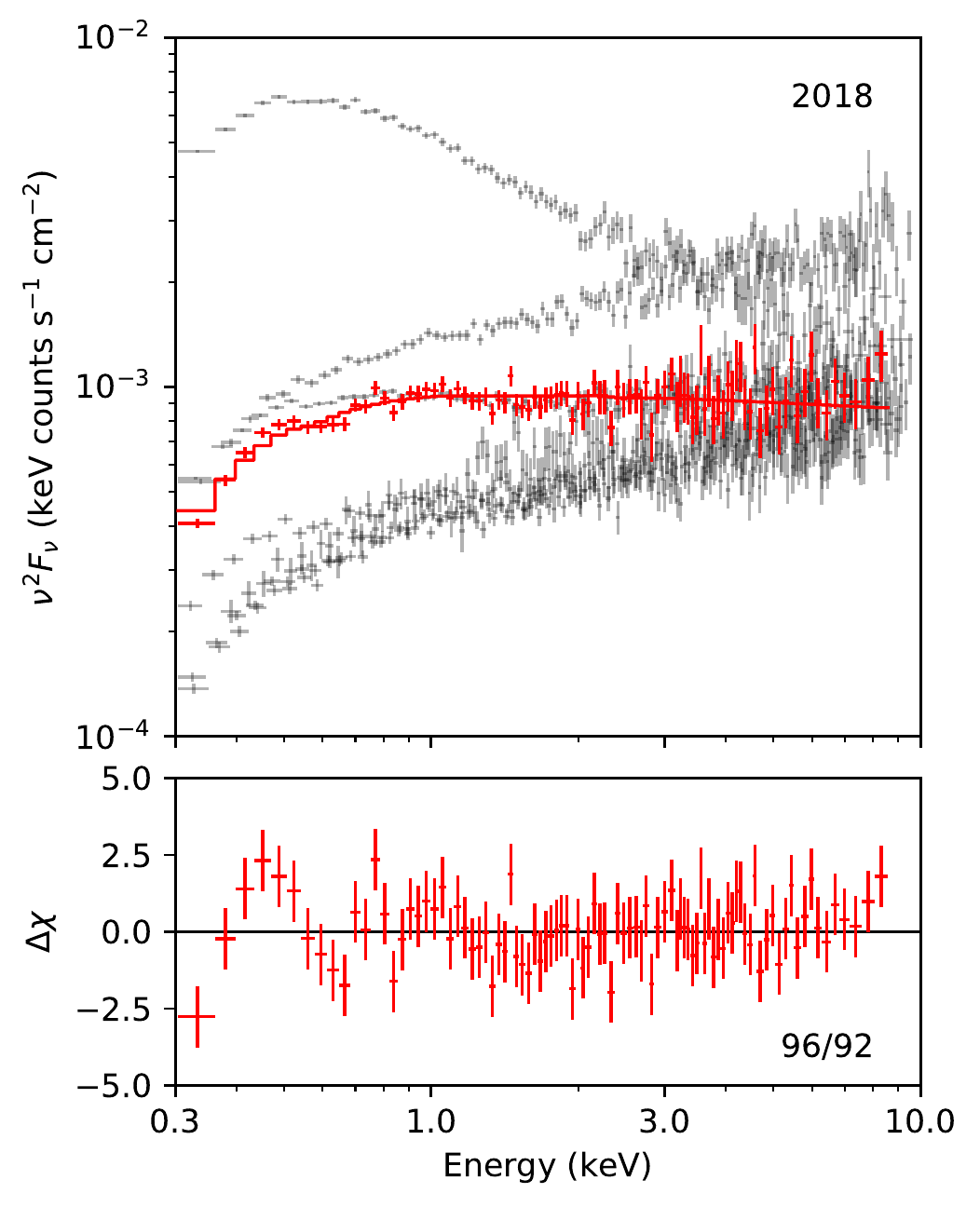}
\includegraphics[width=0.32\linewidth]{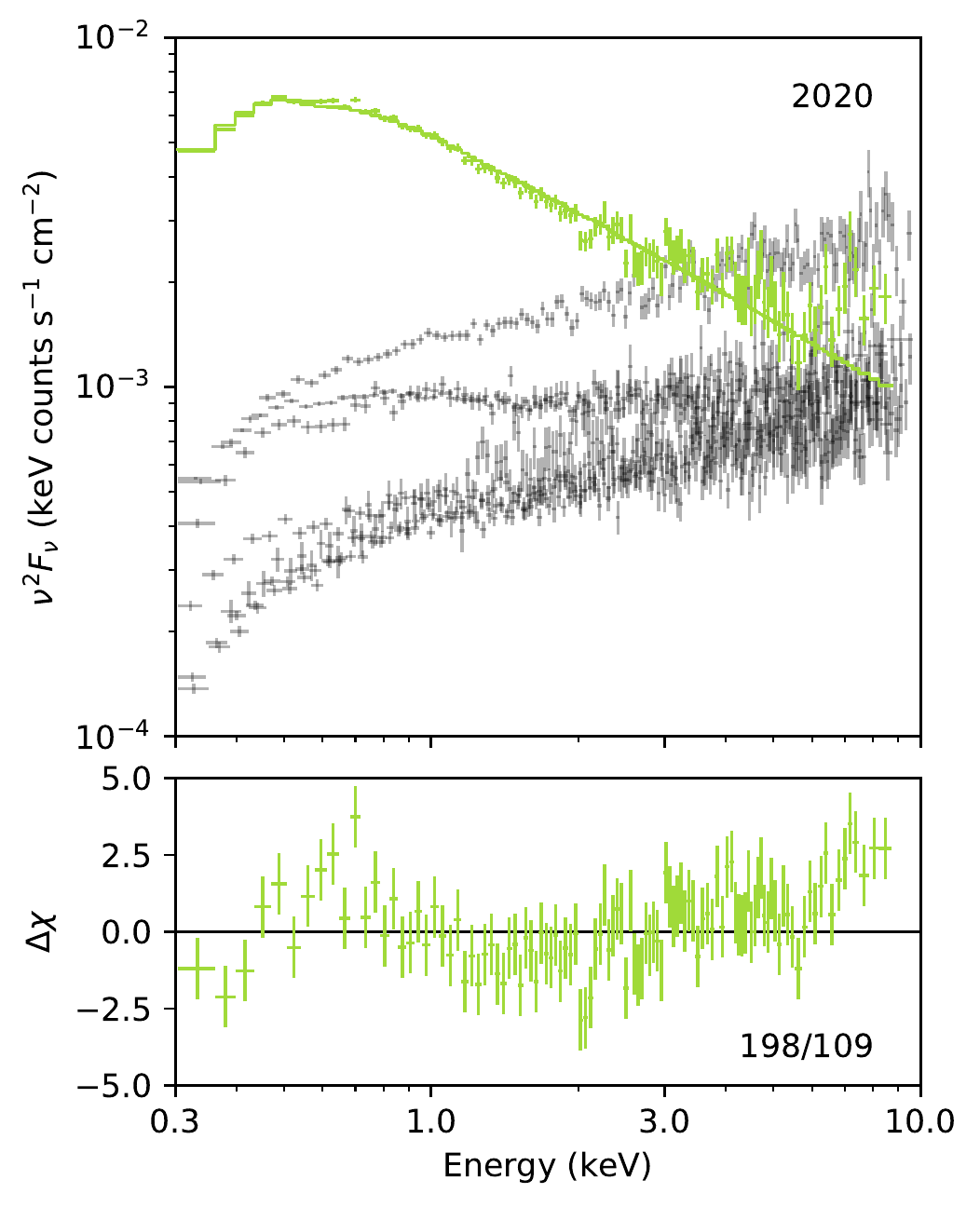}
    \caption{
    Mosaic of single power-law model fits of the XMM-Newton data of OJ 287 and fit residuals in the observer's frame. Each sub-plot corresponds to one XMM-Newton observation of OJ 287 as marked in the upper right corner. The fit statistics are reported in the lower right corner. 
    The residuals are plotted in units of standard deviations, so that structure at low energies is visible and the residuals are not dominated by noisy low-signal bins at high energies.
    }
    \label{fig:XMM-allfits-pl}
\end{figure*}

Results from single power-law fits of all data sets are displayed in Fig. \ref{fig:XMM-allfits-pl}. It is common to display fit residuals of single power-law fits, because these allow to visualize the residuals of this simple standard model, and facilitate comparison of past and future data of OJ 287 and also of other blazar X-ray spectra in a homogeneous way, using the same type of model description. Based on Fig. \ref{fig:XMM-allfits-pl} we proceed with the discussion of X-ray spectral fit results.  

As can be seen from Fig. \ref{fig:XMM-allfits-pl}, two XMM-Newton low-state spectra, taken in 2005a and 2008, are fit well by a single power-law component. Additional components are not required by the fit and are therefore largely unconstrained in a free fit. We return to these low-state spectra below, then taking into account what we have learned from fitting the other spectra. 

\begin{figure*}
\includegraphics[width=0.32\linewidth]{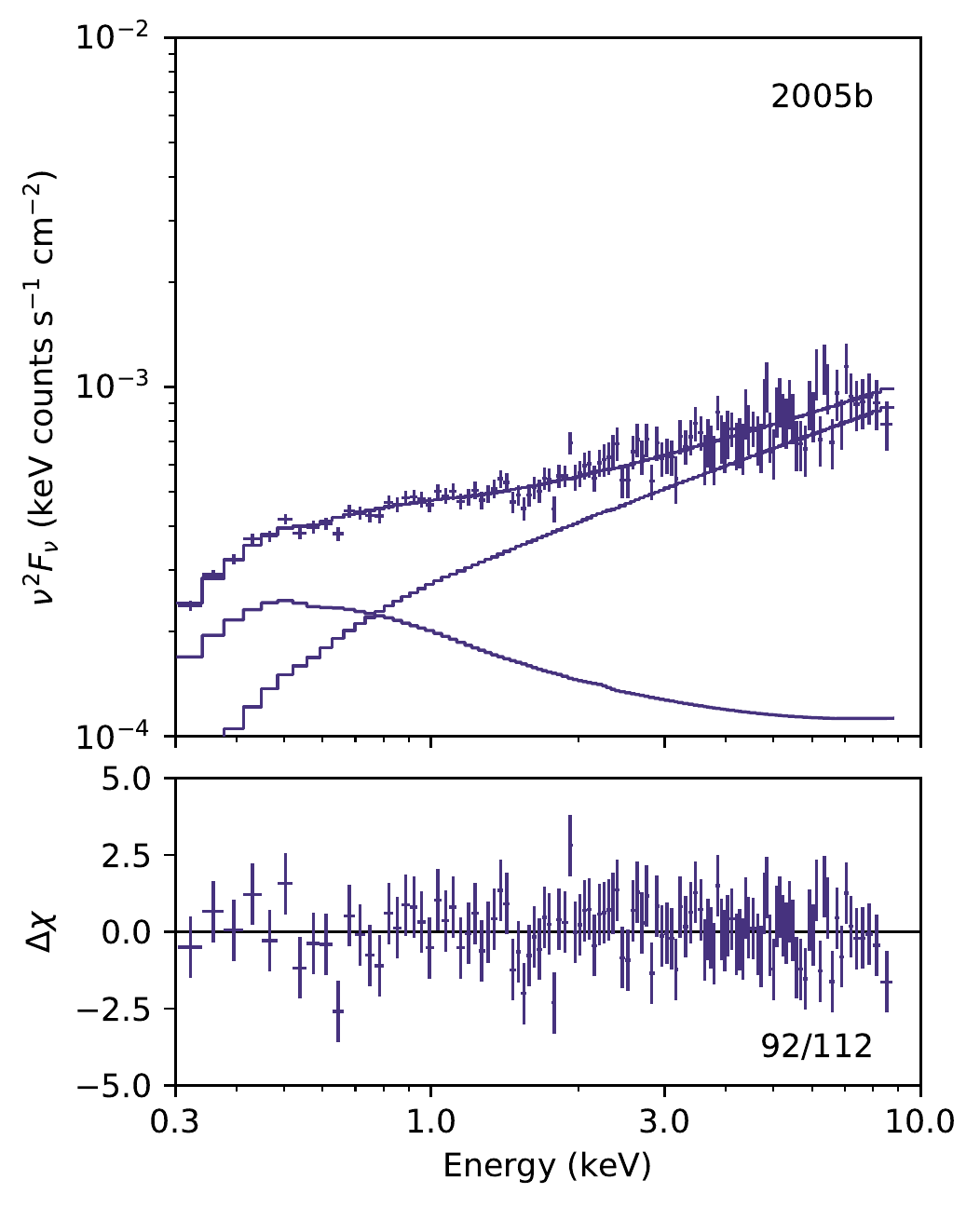}
\includegraphics[width=0.32\linewidth]{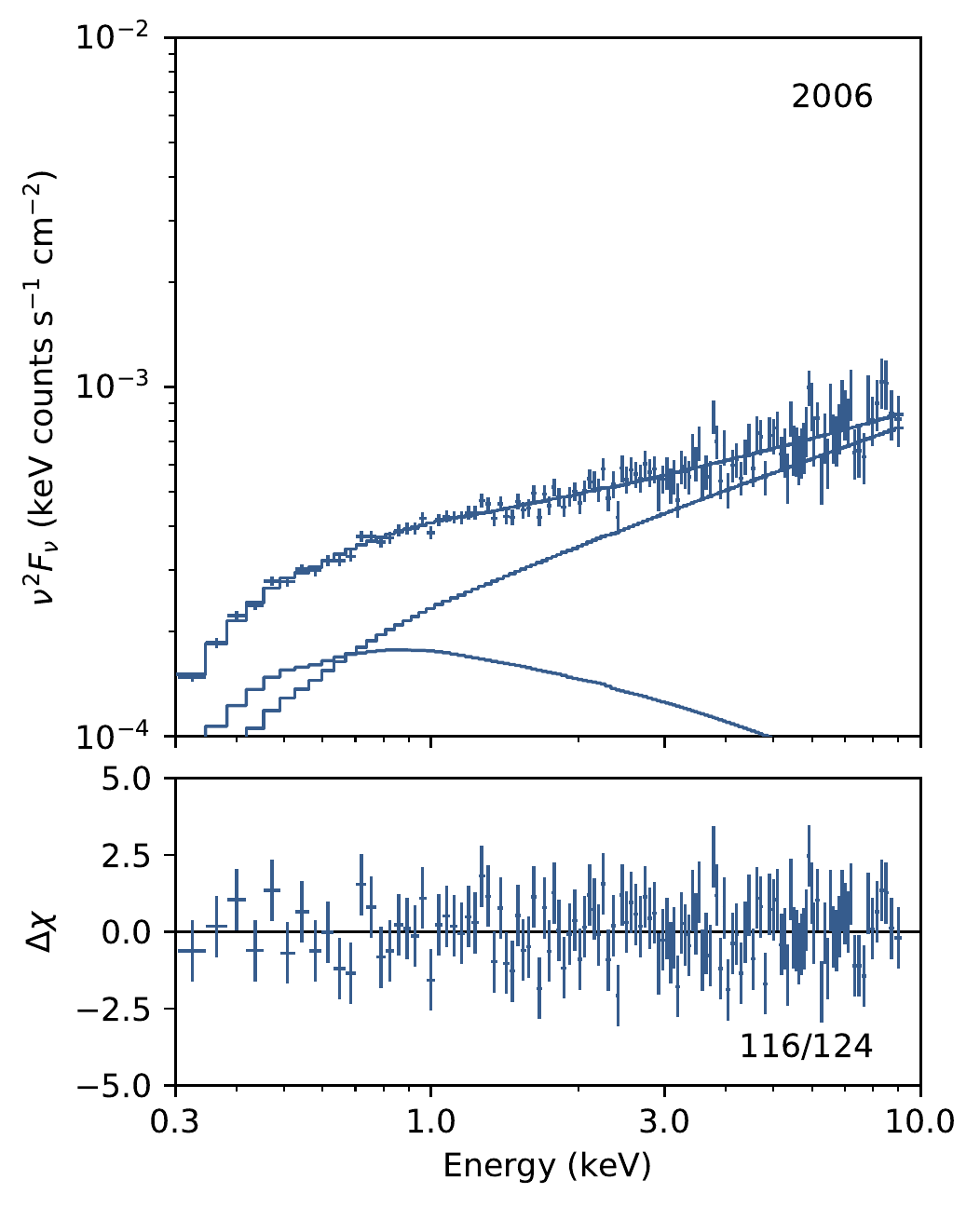}
\includegraphics[width=0.32\linewidth]{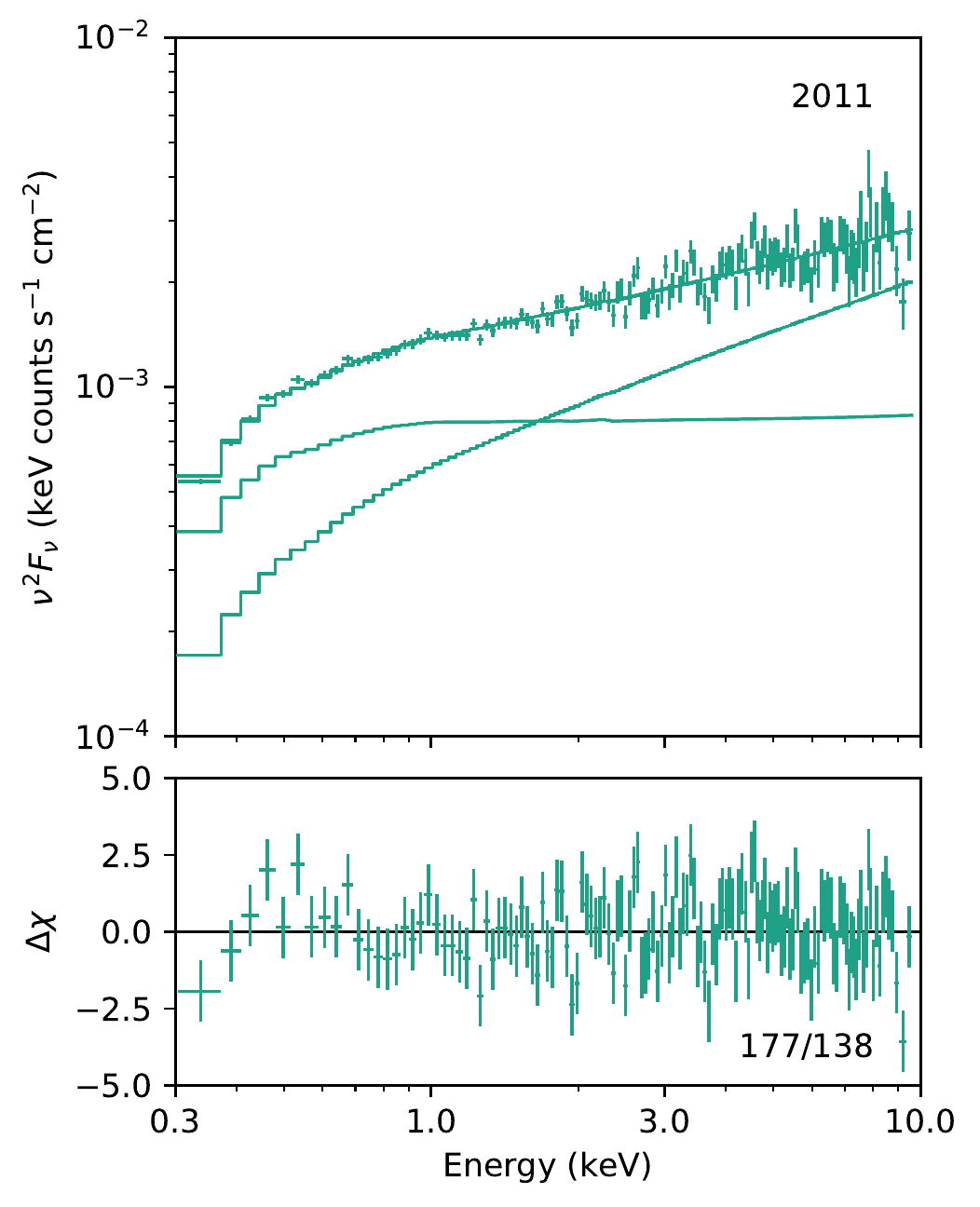}
\includegraphics[width=0.32\linewidth]{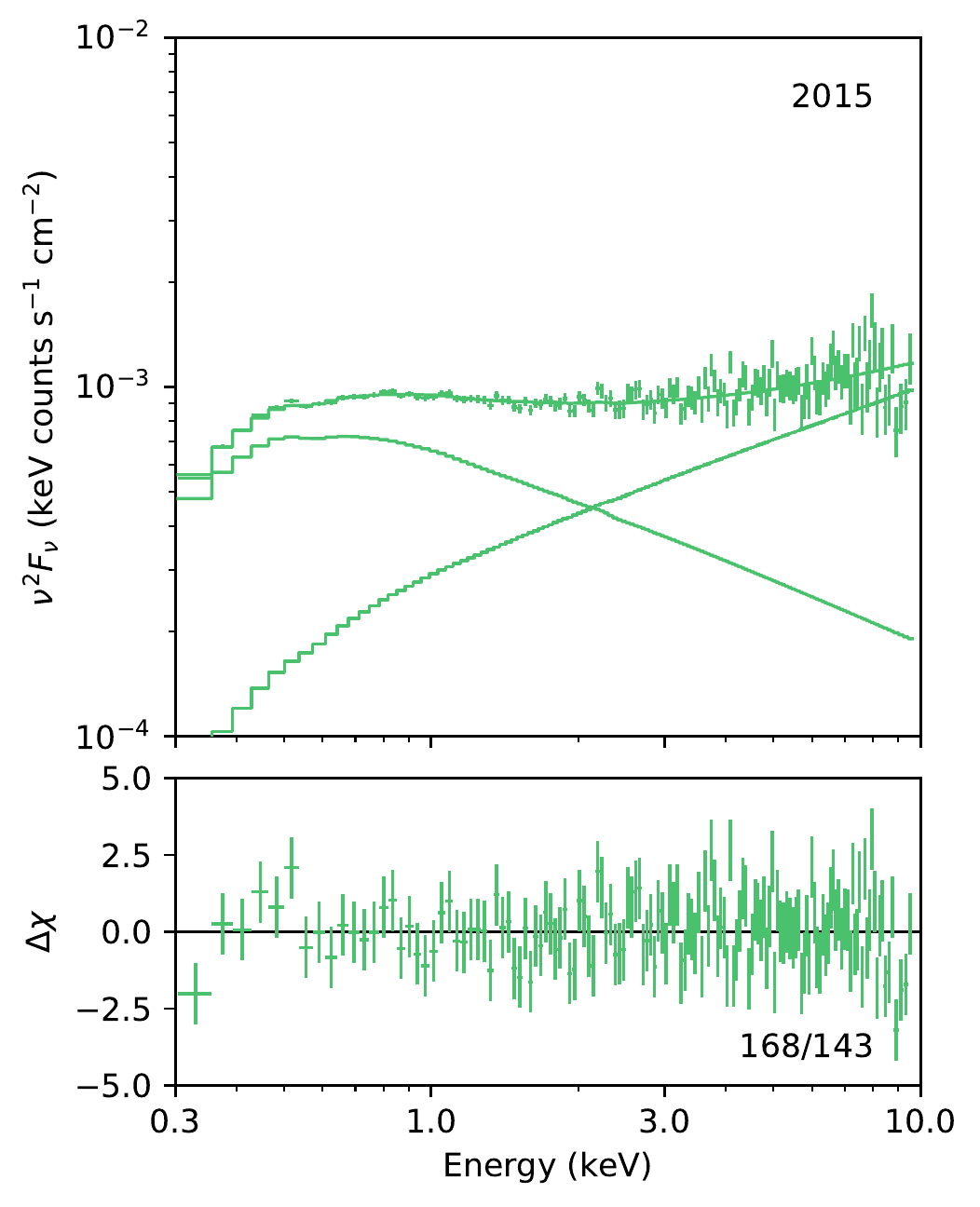}
\includegraphics[width=0.32\linewidth]{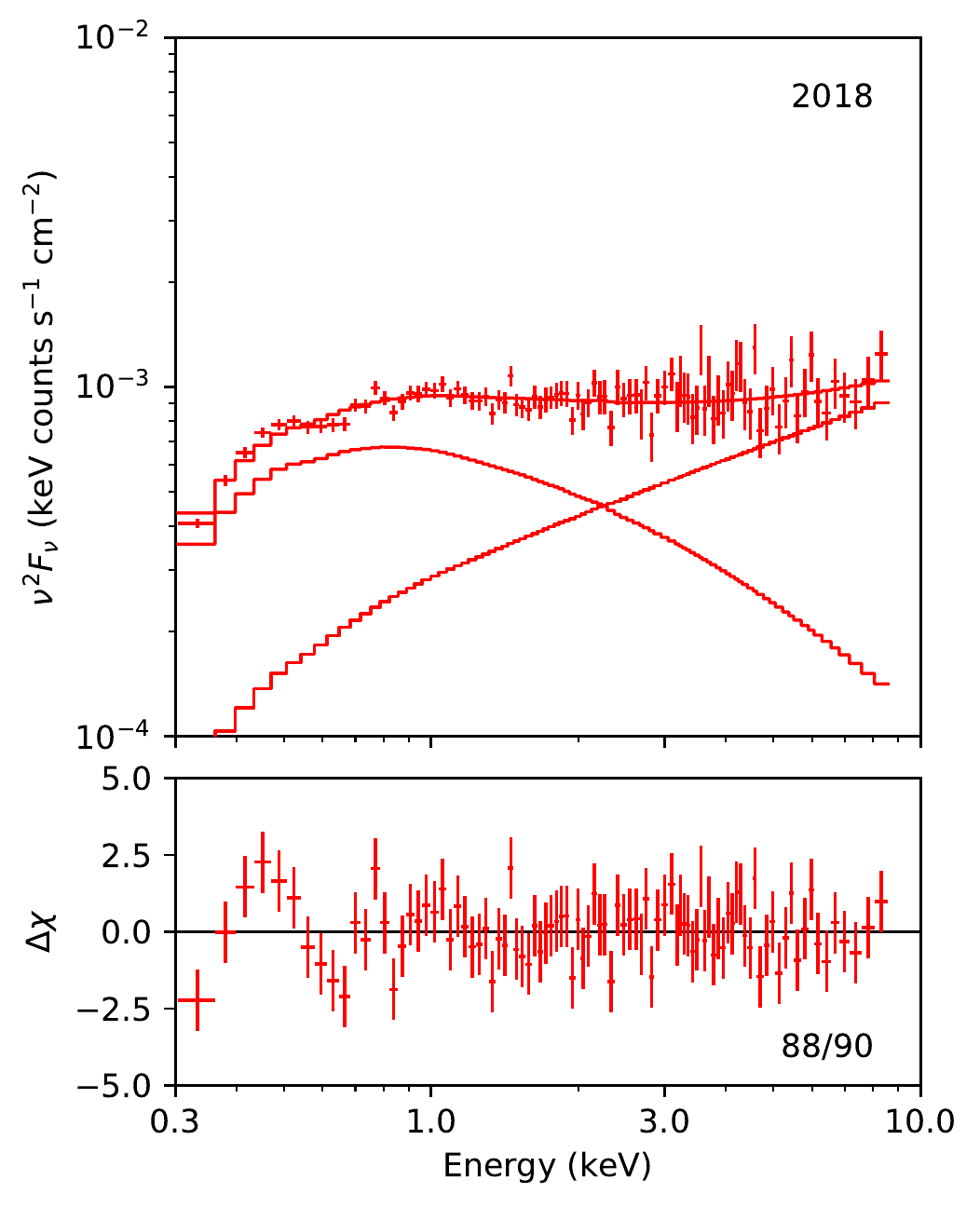}
\includegraphics[width=0.32\linewidth]{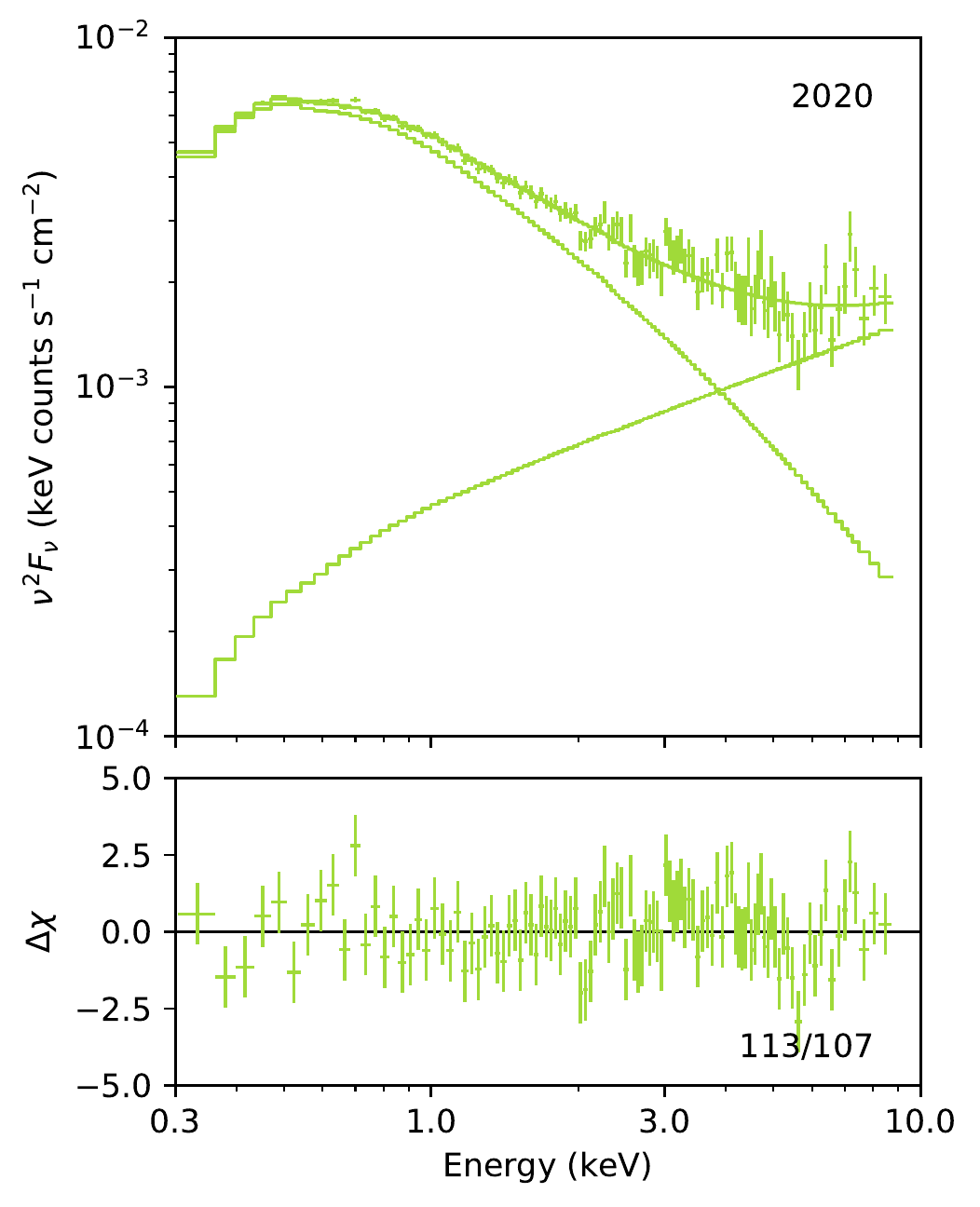}
    \caption{Best-fit logpar plus power law models of those XMM-Newton observations of OJ 287 that require a second component beyond a single power law.
    Each sub-plot corresponds to one XMM-Newton observation of OJ 287. The date is marked in the upper right corner and the fit statistics in the lower right corner. The soft (low-energy) emission component is always the logpar model, while the hard (high-energy) emission component is always the power law.  
    }
    \label{fig:XMM-allfits-bf}
\end{figure*}

The remaining spectra are not well fit by a single power law. The fit quality is not significantly improved when allowing for additional absorption intrinsic to OJ 287, which always turns out to be consistent with zero. 
The spectra require the presence of a second spectral component in addition to a single power law. 
Adding a second power law component or a logarithmic parabolic model fits the spectra well (Tab. \ref{tab:spec-fits} and Fig. \ref{fig:XMM-allfits-bf}), and with the exception of the 2020 data set when OJ 287 was brightest, the two models cannot be distinguished in terms of fit quality. The higher-energy component of the two-component power-law fit is always consistent with a photon index of $\Gamma_{\rm X}=1.5$ within the errors. $\Gamma_{\rm X}$ was therefore fixed to that value in the logpar plus power-law models. 

The steeper power law with photon index in the range $\Gamma_{\rm X} \simeq 2-2.8$ (or the logarithmic parabolic model) accounts for the soft (low-energy) emission component which breaks into a harder power law component with  
$\Gamma_{\rm X} \simeq 1.5$ (in the two-component power-law models) up to 10 keV.
A notable exception is the 2020 outburst spectrum which requires a {\em significantly steeper hard X-ray component}. Fit results of the 2020 data were reported in paper I, and are provided in Tab. \ref{tab:spec-fits} for comparison.  

The phenomenological black body representation of the soft emission component does not provide a successful fit to most of the spectra, including the highest-quality 2020 outburst spectrum which had the strongest soft excess, and we therefore do not discuss it any further, but keep it in Tab. \ref{tab:spec-fits} for comparison purposes. 

Flux changes of OJ 287 are not only driven by the strength of the soft emission component; the hard X-ray emission is variable, too, albeit generally with lower amplitude.
The 2011 XMM-Newton spectrum of OJ 287 stands out as showing the brightest flux above $\sim$5 keV. 

The break energy, where the soft and hard spectral components intersect, is located around 2 keV. It shifts to $<$1 keV in the 2005b and 2006 low-state spectra, and to 4 keV in the 2020 outburst spectrum (Fig. \ref{fig:XMM-allfits-bf}).   

Finally, we return to the low-state spectra of 2005a and 2008, which could be fit with single power laws. Since the two-component power-law fits of the other data sets always require a higher-energy spectral component with $\Gamma_{\rm x} \simeq 1.5$ and a lower-energy soft component, it is plausible to assume that the same holds for the 2005a and 2008 data, where the low photon statistics prevented a two-component spectral decomposition.  As an example, we analyzed the 2008 data further, where $\Gamma_{\rm x}$ = 1.75. 
For a quantitative evaluation of the remaining contribution of a low-energy soft emission component to this spectrum, 
we have enforced a two-component fit with $\Gamma_{\rm X,2}$=1.5. Then, $\Gamma_{\rm X,1} \leq 2$, and up to 50\% of the flux can be contained in such a soft component. 

\subsection{The 2018 data quasi-simultaneous with EHT}

\begin{figure}
\centering
\includegraphics[width=7.0cm]{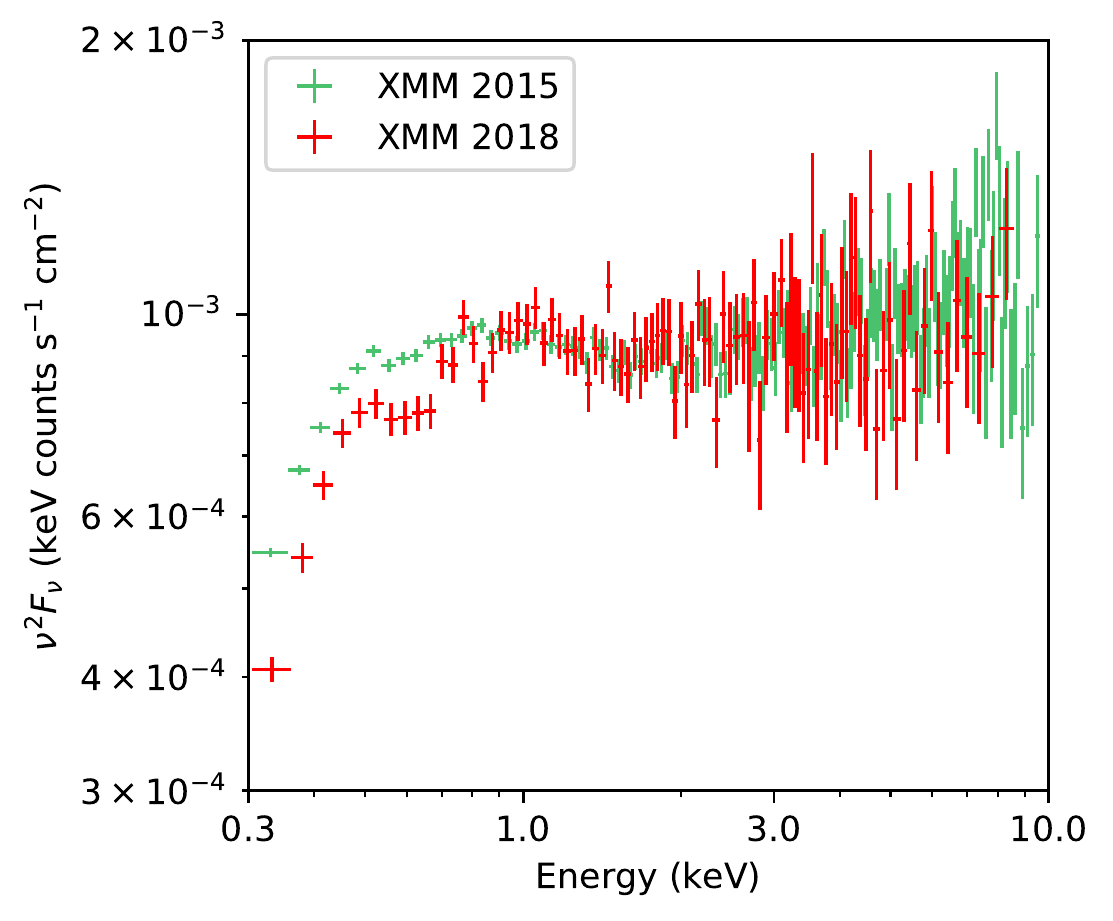}
\includegraphics[width=7.0cm]{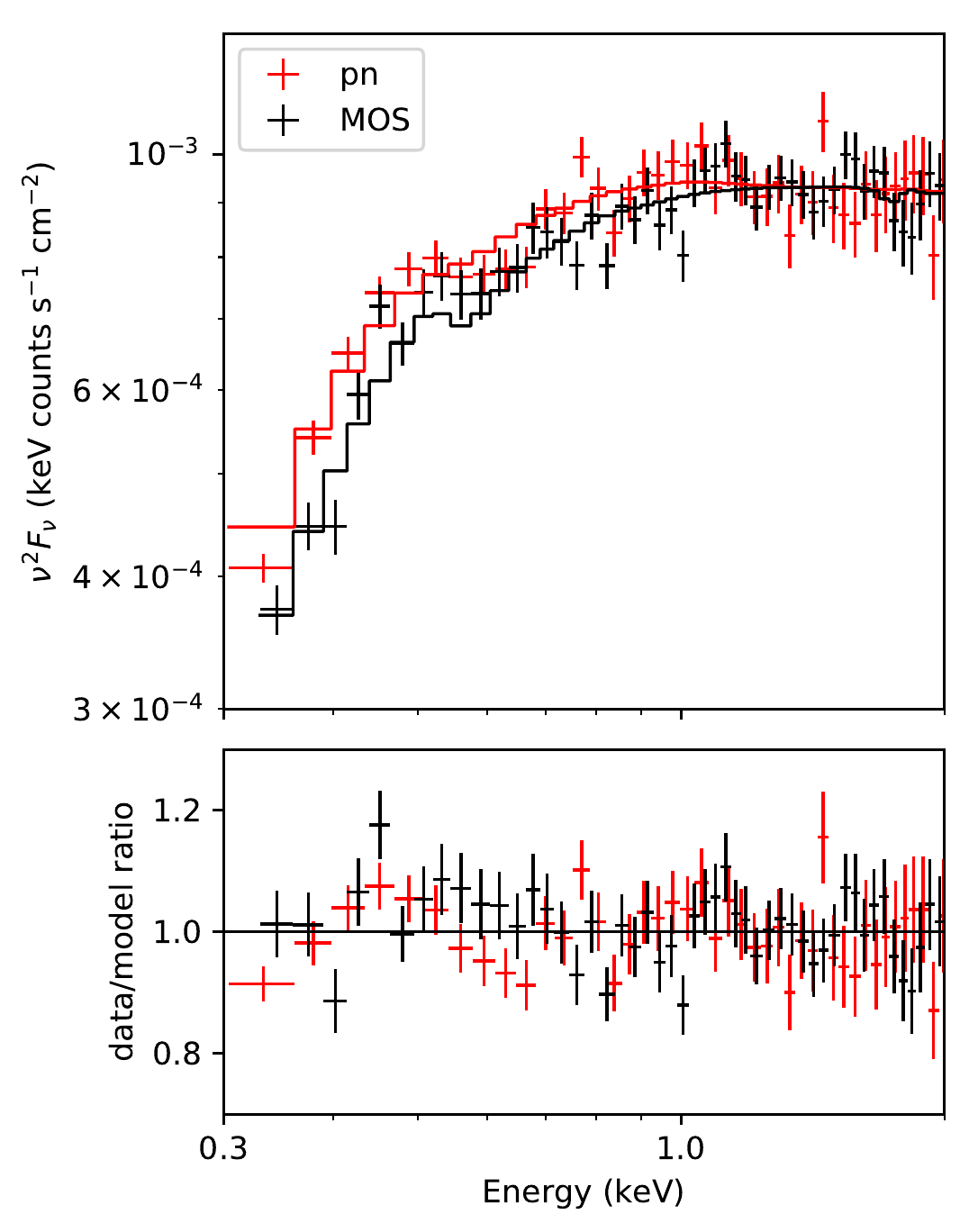}
    \caption{
    2018 XMM-Newton EPIC pn and EPIC MOS spectra of OJ 287 it was at intermediate flux level, observed quasi-simultaneous with EHT. The upper panel shows a comparison between the 2015 and 2018 EPIC pn data, where OJ 287 was overall in a similar state but shows a small low-energy deficiency in flux in 2018.
    The lower panels compare the EPIC pn and EPIC MOS data of 2018. }
    \label{fig:XMM-2018}
\end{figure}

\begin{figure}
    \centering
    \includegraphics[width=\linewidth]{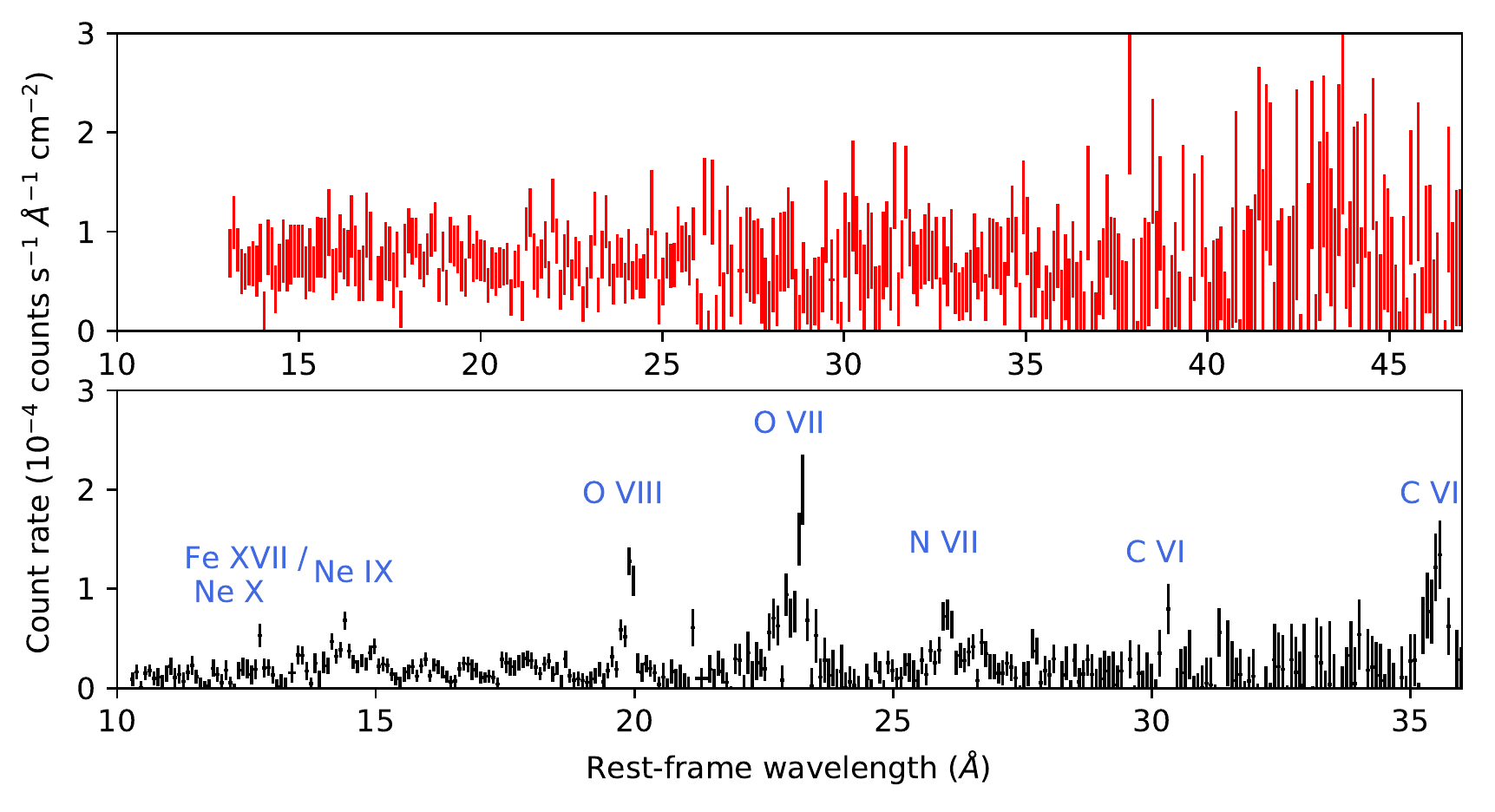}
    \caption{2018 RGS spectrum of OJ 287 (upper panel) in comparison with the RGS spectrum of a Seyfert galaxy (lower panel; both spectra in observer frame) which exhibits strong emission lines from common elements and transitions, labelled in blue \citep[Mrk 335;][]{Parker2019}. No narrow features are detected in the RGS spectrum of OJ 287.
    }
    \label{fig:RGS2018}
\end{figure}

The 2018 observation, quasi-simultaneous with EHT, shows OJ 287 in an intermediate flux state. 
The spectrum shows two components. A hard power-law component with $\Gamma_{\rm x}$=1.5, and a second soft emission component with $\Gamma_{\rm x}$=2. 
Given possible hints of spectral complexity below $\sim$ 1 keV (Figs \ref{fig:XMM-allfits-pl}, \ref{fig:XMM-2018}),
the 2018 data were carefully compared with the 2015 exposure where OJ 287 was in a similar flux state.
Further, the pn and MOS spectra of the 2018 observation were inspected and compared, revealing that fine structures around 0.8 kev in the pn data are not present in MOS and are therefore not statistically significant (Fig. \ref{fig:XMM-2018}). 
Independently, RGS data were inspected for the presence of strong narrow emission or absorption features. None were individually identified (Fig. \ref{fig:RGS2018}). 
When an ionized absorber component is added to the spectral fit using the \textsc{xabs} model \citep{Steenbrugge03} the quality of the spectral fit is not significantly improved.  
Most likely, the small deficit of photons below 1 keV in 2018 as compared to 2015 is due to a slightly fainter soft emission component in 2018. 

\subsection{Short timescale variability}

The longest observation of OJ 287 was carried out in 2015 with a duration of 121 ks. Since OJ 287 is highly variable on timescales of days and longer, this light curve was inspected for shorter timescale variability and flaring. Since $\sim$50\% of the observation was affected by episodes of background flaring which dominate
the emission at $>$2 keV, only the full (0.3--1.5) keV light curve was analyzed.
Some limited variability is visible on relatively long timescales (10s of ks), with no clear evidence for rapid variability. We calculate the fractional root mean square (RMS) variability amplitude $F_\mathrm{var}$ \citep[e.g.][]{Vaughan2003} for this light curve, finding a small but significant (dimensionless) value of 
$F_\mathrm{var,s}$=$0.033\pm0.002$. 
Since the hard X-ray band is affected by multiple episodes of background flaring (Sect. 3.1), we only analyzed the first good contiguous 40 ks of the hard-band light curve and find $F_\mathrm{var,h}$=$0.04\pm0.01$. 

\setlength{\tabcolsep}{3pt}
\begin{table*}
 \scriptsize
	\centering
	\caption{
	Spectral fit results of all XMM-Newton observations of OJ 287. Fits to the 2020 data were already reported in paper I. Representative models are repeated here for comparison with the earlier observations. 
	Neutral absorption of the Galactic value $N_{\rm H, Gal}$ is included in all fits. 
	Parameters and abbreviations are as follows: (1) Models: pl = power law, bbdy = black body, logpar = logarithmic parabolic model; 
	(2) absorption column density in units of 10$^{20}$ cm$^{-2}$; (3) power-law photon index; (4) unabsorbed power-law flux ($f_1$, $f_2$) or logpar flux ($f_{\rm lp}$) from (0.5--10) keV 
	in units of 10$^{-12}$ erg\,s$^{-1}$\,cm$^{-2}$ (rest frame); (5) black body temperature $kT_{\rm BB}$ in units of keV; (6) (0.5--10) keV black body flux in units of 10$^{-12}$ erg\,s$^{-1}$\,cm$^{-2}$ (rest frame);  
	(7) curvature parameter $\beta$ and (8) spectral index $\alpha$ of the logpar model; (9) goodness of fit $\chi^2_{}$ and number of degrees of freedom. When no errors are reported, the quantity was fixed. The spectral fits of the NuSTAR data, taken from paper I, are shown for comparison.
	For NuSTAR data, the pl flux is given from 3--50 keV. 
	}
	\label{tab:spec-fits}
	\begin{tabular}{lccccccccccc}
		\hline
		model & $N_{\rm H}$ & $\Gamma_{\rm 1, soft}$ & $f_1$ & $\Gamma_{\rm 2, hard}$ & $f_2$ & $kT_{\rm BB}$ & $f_\mathrm{BB}$ & $\beta$ & $\alpha$ & $f_\mathrm{lp}$ & $\chi{^2}$/$n{_\mathrm{dof}}$ \\
		(1) & (2) & (3) & (4) & (3) & (4) & (5) & (6) & (7) & (8) & (4) & (9) \\
		\hline
		\multicolumn{5}{l}{2020, highest state}\\
		\hline
		pl & 2.49 & $2.82\pm0.01$ & $38.5\pm0.1$& - & - & - & - & -& - & - & 476/284 \\
		bbdy + pl & 2.49 & $2.70\pm0.01$ 
		& $35.2\pm0.3$ & - & - & $0.152\pm0.003$ &  $3.2\pm0.4$ &-&-& -& 343/282 \\
		logpar + pl & 2.49 & - & - & 2.2 & $17.3\pm0.6$ &- &- & $0.8\pm0.1$ & $3.22 \pm0.03$ &  $21.5\pm0.6$$^{1}$ & 286/282 \\
pl (NuSTAR) & 2.49 & - & - & $2.36\pm0.06$ & $6.1\pm0.2$ & -&-&-&-&-& 74/74 \\
pl (\scriptsize{NuSTAR, $>$10 keV}) & 2.49 & - & - & $2.2\pm0.2$  & - & - & - & - & - & - &  \\
		\hline

\multicolumn{5}{l}{2018, simultaneous with EHT}\\
		\hline
pl & 2.49 & $2.08\pm0.01$ &$8.10\pm0.07$ &- &- &- &- &- &- &- & 96/92\\ 
pl, $N_{\rm H}$ free & $<0.7$ + 2.49 & $2.08^{+0.03}_{-0.02}$ &$8.1\pm0.1$ &- &- &- &- &- &- &- & 96/91\\

bbdy + pl & 2.49 & $2.02\pm0.03$ & $7.8\pm0.1$ &- &- & $0.20\pm0.02$& $0.3\pm0.1$ &- &- &- & 89/90\\ 

pl + pl & 2.49 & $2.12_{-0.01}^{+0.02}$ & $7.2\pm0.6$& $<1.55$ &$0.9_\pm0.6$ &- &- &- &- &- & 93/90\\

logpar + pl & 2.49 & - & - & 1.5 & $3.7_{-0.6}^{+0.4}$ &- &- & $0.41\pm0.16$ & $2.30\pm0.05$ & $4.5_{-0.4}^{+0.6}$ & 87/90 \\
		\hline

2015 & & & & &  &  & & & & & \\
        \hline
pl & 2.49 & $2.188\pm0.005$ & $8.25\pm0.03$&- &- &- &- &- &- &- & 734/145 \\

pl, $N_{\rm H}$ free & $<0.01$ + 2.49 & $2.19\pm0.01$ & $8.25\pm0.08$ & - &- &- &- &- &- &- & 734/144\\

bbdy + pl & 2.49 & $2.00\pm0.01$ & $7.91\pm0.04$ &- &- & $0.130\pm0.003$& $0.69\pm0.04$ & -& -& -& 222/143\\ 

pl + pl & 2.49 & $2.41_{-0.01}^{+0.02}$ & $4.7_{-0.5}^{+0.3}$ & $1.54\pm0.08$ &$3.9_{-0.3}^{+0.6}$ &- &- &- &- &- & 168/143\\

logpar + pl & 2.49 & - & - & 1.5 & $3.7_{-0.4}^{+0.3}$ &- &- & $0.013\pm0.008$ & $2.58_{-0.04}^{+0.03}$ & $4.90_{-0.0.3}^{+0.4}$ & 167/143 \\
		\hline

2011 & & & & &  &  & & & & & \\
        \hline
pl & 2.49 & $1.799\pm0.006$ &$14.39\pm0.08$ &- &- &- &- &- &- &- & 233/140\\ 

bbdy + pl & 2.49 & $1.72\pm0.01$ &$14.2\pm0.1$ &- &- &$0.15\pm0.01$ & $0.46\pm0.07$ &- &- &- & 166/138\\

pl + pl & 2.49 & $2.4\pm0.3$ & $2.9_{-1.6}^{+2.9}$ & $1.6\pm0.1$ & $11.7_{-2.9}^{+1.6}$&- &- &- &- &- & 177/138\\

logpar + pl & 2.49 & - & - & 1.5 & $7.5_{-5.6}^{+2.2}$ &- &- & $7.1_{-2.2}^{+5.6}$ & $1.6\pm0.1$ & $4.1_{-1.4}^{+3.8}$ & 177/138 \\
		\hline
		
2008 & & & & &  &  & & & & & \\
        \hline
pl & 2.49 & $1.752\pm0.009$& $4.34\pm 0.03$ &- &- &- &- &- &- &- & 121/121\\ 
pl + pl & 2.49 & $<2$ & $ 2.0\pm0.1 $ & 1.5 & $ 2.3\pm0.1 $ &- &- &- &- &- & 121/120\\
		\hline
		
2006 & & & & &  &  & & & & & \\
        \hline
pl & 2.49 & $1.80\pm0.01$&$4.23\pm 0.03$ &- &- &- &- &- &- &- &149/126 \\ 

bbdy + pl & 2.49 &$1.74\pm0.02$ & $4.19\pm0.04$ &- &- &$0.14\pm0.02$ & $0.10\pm0.02$ &- &- &- &123/124 \\ 

pl + pl & 2.49 & $1.96_{-0.05}^{+0.29}$ & $1.7_{-0.9}^{+0.0.2}$ & $<1.6$ & $2.6_{-0.2}^{+0.8}$ &- &- &- &- &- & 116/124\\

logpar + pl & 2.49 & - & - & 1.5 & $<5.5$ &- &- & $0.17_{-0.37}^{+0.24}$ & $2.2_{-0.3}^{+0.1}$ & $1.3_{-0.2}^{+4.2}$ & 116/124 \\
		\hline

2005-11 (2005b) & & & & &  &  & & & & & \\
        \hline
pl & 2.49 & $1.92\pm0.01$ & $4.85\pm0.04$ &- &- &- &- &- &- &- & 308/114\\ 

bbdy + pl & 2.49 &$1.75\pm0.02$ & $4.83\pm0.05$ &- &- &$0.106\pm0.007$ & $0.31\pm0.03$ &- &- &- & 99/112\\ 

pl + pl & 2.49 & $>2.95$ & $1.12_{-0.06}^{+0.08}$ & $1.54\pm0.03$ & $4.04\pm 0.07$ &- &- &- &- &- & 92/112\\

logpar + pl & 2.49 & - & - & 1.5 & $3.5_{-0.5}^{+0.3}$ &- &- & $-0.4_{-0.1}^{+0.2}$ & $2.7\pm0.1$ & $1.6_{-0.3}^{+0.5}$ & 92/112 \\
		\hline

2005-04 (2005a) & & & & &  &  & & & & & \\
        \hline
pl & 2.49 & $1.60\pm0.02$ &$5.9\pm 0.1$ &- &- &- &- &- &- &- &102/98 \\ 
		\hline
	\end{tabular}

$^1$ Note: in paper I (Tab. 2), the flux entry for the logpar component of this model was a typo, now corrected.    
	
\end{table*}

\section{Discussion} 

The spectral energy distribution (SED) of blazars shows two emission maxima \citep[for reviews see:][]{Marscher2009, Ghisellini2015}.
One at low energies peaking between the sub-mm and EUV, sometimes extending into the X-ray regime, and explained as synchrotron radiation of the jet electrons. And a second maximum in the hard X-ray/$\gamma$-ray regime, usually explained as inverse Compton (IC) radiation from photons located either within the jet  (synchrotron-self-Compton process; SSC) or from external seed photons emitted by the broad-line region or torus (external comptonization; EC).
Other radiation mechanisms might contribute as well. 
In hadronic models the high-energy emission is dominated by ultra-relativistic protons \citep{Boettcher2019}. 
The SED-type of OJ 287 was classified as
LSP (low synchrotron peak frequency; $<10^{14}$ Hz; \citet{Abdo2010}) 
or ISP (intermediate synchrotron peak frequency; between 10$^{14}$ and 10$^{15}$ Hz; \citet{Ackermann2011}). An additional thermal component in its broad-band SED was identified at epochs close to the impact flare predicted by the SMBBH model \citep[][]{Ciprini2007, Ciprini2018, Valtonen2012}.  

Previous observations suggested that variability in X-rays of OJ 287 is driven
primarily by a mix of the SSC component and synchrotron emission.
However, most previous missions did not allow multi-component spectral fits, 
and during the 
2020 XMM-NuSTAR high-state (paper I), the high-energy power-law component was steeper than expected for pure IC emission models which produce hard spectra with $\Gamma_{\rm X} \approx 1.5$ \citep{Kubo1998, Isobe2001}. We discuss implications from the XMM-Newton spectral fits below.

\subsection{Softer-when-brighter variability pattern} 

OJ 287 exhibits a softer-when-brigther variability pattern in our multi-year Swift light curve \citep[Fig. \ref{fig:Swift-HR}; see also][]{Komossa2017},  previously sporadically seen on long timescales when combining a few Einstein, EXOSAT, ROSAT, and ASCA data \citep{Isobe2001, Urry1996},
but absent at another epoch \citep{Seta2009, Siejkowski2017}.  

The Swift XRT light curve, containing 681 data points obtained with the same instrument, for the first time puts the softness-brightness behaviour on a firm statistical basis and also traces short timescales of days, weeks, and months, and reveals details of the correlation. 
While the short snapshot observations with Swift do not contain enough photons to permit multi-component spectral decompositions, the much deeper observations with the more sensitive detectors aboard XMM-Newton allowed us to decompose the spectra into multiple components spanning 15 years of observations. These components will be discussed in the next sections. 
In particular, the spectral analysis of the XMM-Newton observation of 2011 has shown that variability of the high-energy spectral component (Fig. \ref{fig:XMM-allobs}) contributes to the scatter in the softer-when-brighter variability at low countrates ($<0.5$ cts s$^{-1}$). This epoch of observation coincides with $\gamma$-ray flaring activity of OJ 287 \citep[e.g.,][]{Escande2011, Hodgson2017, Goyal2018} of which we see the low-energy tail in the XMM-Newton band.

\subsection{Synchrotron component(s) at high-state} 
Which mechanism drives the super-soft component of the X-ray spectrum in high-states: accretion or jet (synchrotron) activity ? We can safely establish that the 2020 outburst is powered by jet emission, because (1) the rapid X-ray variability is inconsistent with the large SMBH mass of the primary (faster than the light crossing time at the last stable orbit of the accretion disk; paper I), (2) the accompanying radio flare in 2020 \citep{Komossa2020c} implies jet emission in the radio regime, (3) the high optical polarization \citep{Zola2020} implies jet emission in the optical, too, and (4) the correlated (radio-)optical--UV--X-ray emission then implies jet emission in all bands.  

The second large X-ray outburst of OJ 287 was in late 2016 -- early 2017  \citep[our Fig. \ref{fig:lc-Swift};][]{Komossa2017}, and given its similarities with the 2020 outburst and additional arguments, we argue that it was jet-powered, too. It shows the very same softer-when-brighter variability pattern (established in our Fig. \ref{fig:Swift-HR}), it was accompanied by a radio outburst \citep{Myserlis2018, Lee2020} and by VHE emission \citep{O'Brien2017}, and the radio polarization was very high during a contemporaneous ALMA observation (Goddi et al. 2021). 

Interestingly, during the 2020 outburst, where we obtained a dedicated XMM-Newton and NuSTAR observation at the peak of the outburst, the X-ray spectrum is more complicated than a superposition of a low-energy synchrotron and high-energy IC component. Besides the very soft emission component well described by a logpar model, the power-law component which extends well above 10 keV is steeper than expected from IC emission alone and we therefore consider it to be a third spectral component. It may plausibly indicate a mix of a flat IC component with a second independent synchrotron component extending well above 10 keV only appearing at outburst, or alternatively could represent temporary enhanced emission from an accretion disk corona.
While we resolve the emitting components spectroscopically, high-resolution radio observations hold the promise to locate the synchrotron emission components spatially. No EHT observations were carried out in 2020, but they are expected to continue in upcoming years. The unusual second 2020 (synchrotron or coronal) high-state component adds to making OJ 287 an interesting target in upcoming EHT campaigns. 

 The rapid variability of the multiple synchrotron components 
 is reminiscent of the erratic jet model of \citet{Agudo2012}, who involve multiple injections from a turbulent and/or clumpy disk to explain
 (short-term) erratic wobbling of the inner jet observed in VLBA data of OJ 287 on the order of months.    
 The frequent secondary impacts on the accretion disk around the primary predicted by the binary SMBH model of OJ 287 might contribute to enhancing inhomogeneities/turbulence in the accretion disk of OJ 287. Additionally, \citet{ValtonenPihajoki2013} and \citet{Dey2021} explain the long-term systematic trends in the wobble by variations in the initial jet launching angle in the binary system. 

\subsection{Low-state emission} 
During X-ray low-states of OJ 287, the soft (synchrotron) component is much fainter or entirely absent, and the XMM-Newton X-ray spectrum is well fit by a flat power-law component as flat as $\Gamma_{\rm x} \approx 1.5$, indicating IC emission (in the framework of leptonic jet models). Similarly, the flattest X-ray states of OJ 287 
detected with Swift are well described by a single power law with photon index 1.5. 

We use the 2008 low-state of OJ 287 measured with XMM-Newton to estimate the low-state {\em isotropic} X-ray luminosity, and take it as a strict {\em upper limit} on any long-lived accretion disk contribution to the X-ray spectrum of OJ 287. 
With $L_{\rm x, iso} = 1.3 \times 10^{45}$ erg\,s$^{-1}$  
this gives $L_{\rm x}/L_{\rm Edd}  \leq 5.6 \times 10^{-4} $ ($M_{\rm BH, primary}=1.8
\times 10^{10}$ M$_\odot$) or 
$L_{\rm x}/L_{\rm Edd} \leq 6.7 \times 10^{-2} $ ($M_{\rm BH, secondary}=1.5 \times 10^{8}$ M$_\odot$)
for any accretion disk contribution in X-rays. 

\subsection{Binary black hole model} 

In the context of the SMBBH model of OJ 287 \citep[recently reviewed by][]{Dey2019},
the main site of multi-wavelength emission is still the jet of the primary SMBH, but there are also episodes where the thermal bremsstrahlung emission from the secondary's impact on the accretion disk around the primary \citep{Ivanov1998} dominates the spectrum in the IR to UV bands \citep[e.g.,][]{Laine2020}. 
In addition, there is delayed temporary enhanced accretion on the primary (triggering new jet activity), following the disk impact events. 
Even though the observed low-state X-rays of OJ 287 are dominated by jet activity and consistent with IC emission, 
we used the 
low-state X-ray flux to estimate an upper limit on any long-lived emission from the accretion disk in X-rays (Sect. 4.3).
Assuming a factor-of-10 bolometric correction \citep{Elvis1994}, this then translates into a very low
$L/L_{\rm Edd}  \leq 5.6 \times 10^{-3}$ for $M_{\rm BH, primary}=1.8 \times 10^{10}$ M$_\odot$.
However, this estimate is uncertain for several reasons.
For instance, we do not know the disk emission efficiency $\eta$, and it is possible that OJ 287 accretes near Eddington, but the gas emissivity is low. 
Second, as the SMBH mass increases, we expect to see less (multi-temperature black body) emission from the actual accretion disk extending into the soft X-ray regime \citep{Done2012}, but still expect the presence of hard emission from the corona above the disk (and reprocessing signatures at soft X-rays).  Bolometric corrections are therefore more uncertain at the highest SMBH masses. 
They also depend on accretion rate \citep[]{Vasudevan2009,Grupe2010}. The mass of the primary SMBH itself is very well constrained to be above 10$^{10}$ M$_{\odot}$ in the binary SMBH model \citep{Pietila1998,Dey2018}.
Since previous modelling of OJ 287 in the context of the binary SMBH model requires an accretion rate of $\sim10\%$ Eddington
\citep{Valtonen2019}, within this model the X-ray observations then imply that the disk corona of OJ 287 is under-luminous in X-rays. It is not expected that the corona is completely absent, as the model assumes a highly magnetized accretion disk. 

Swift recorded two major X-ray--optical outbursts of OJ 287 in 2016-17 and 2020 (Fig. \ref{fig:lc-Swift}). 
Both outbursts are driven by synchrotron emission. Based on predictions \citep{Sundelius1997} of the SMBBH model of OJ 287, \citep{Komossa2020a} concluded that the 2020 outburst is consistent with an after-flare which followed the 2019 impact flare of OJ 287.
It is tempting to speculate that, similarly, the 2016-17 X-ray outburst was the after-flare of the 2015 impact flare.
Alternatively, it could represent jet flaring activity unrelated to the binary. 
In the after-flare interpretation, the larger time delay then arises because the secondary's disk impact site was further out in 2015 at larger distance from the primary \citep{Dey2018, Valtonen2019}. 
The after-flare interpretation was already suggested based on {\em optical} high-state observations and high levels of polarization lasting until the end of 2016 \citep{Valtonen2017}. The X-ray high-state continues into the first months of 2017 and reaches its maximum in early 2017. 

With highest-resolution radio observations including EHT, we may be able to resolve the synchrotron outburst components in the radio regime, and locate the outburst emission within the nuclear region. Especially, if the new jet component is powered by a (binary-triggered) new jet event, it will be located in the core; if it is rather an unrelated jet-ISM shock, it could be significantly off-nuclear.

\section{Summary and conclusions}

Based on 1.5 decades of XMM-Newton spectroscopy and Swift monitoring of OJ 287 (densely since 2015 in our dedicated MOMO program aimed at testing facets of the binary SMBH model and studying disk and jet physics of this nearby, bright blazar), the following conclusions are reached:  

\begin{itemize}

\item 
OJ 287 displays a strong softer-when-brighter spectral variability pattern, clearly established on the basis of $\sim$700 Swift observations.  

\item The spectral variability between 0.3--10 keV is extreme. It encompasses all spectral states observed in blazars in that band, from ultra-steep to very flat ($\Gamma_{\rm x}$ = 2.8 -- 1.5). 

\item With XMM-Newton and NuSTAR, we distinguish between three different spectral components of OJ 287: 
(1) The first spectral state is a low/hard state with photon index $\Gamma_{\rm x} \simeq 1.5$, dominated by inverse-Compton emission. (2) The second spectral state is a high/soft state dominated by synchrotron emission well described by a logarithmic parabolic power-law model or with an equivalent single photon index as steep as $\Gamma_{\rm x} \sim 2.8$ in outburst. The softer-when-brighter variability behaviour is explained by the systematic increase of the synchrotron component as OJ 287 brightens.
(3) An additional spectral component is detected in outburst in 2020, with a relatively steep power law of index $2.2\pm{0.2}$ beyond 10 keV observed with NuSTAR and first identified in paper I, likely representing a mix of IC with a second independent synchrotron component. 

\item OJ 287 is highly variable, with rapid flaring on the timescale of days, faster than the light-crossing time near the last stable orbit (paper I). However, the fractional variability amplitude on shorter timescales during a 120 ks XMM-Newton long-look at intermediate flux levels, $F_{\rm var}=0.033\pm0.002$, is small.

\item The 2018 XMM-Newton spectrum which was obtained quasi-simultaneous with EHT shows OJ 287 in a typical intermediate-flux state and is well described by a two-component model, involving a hard power law component with
$\Gamma_{\rm x} = 1.5$, and a second soft emission component in form of a logarithmic parabolic model or a single power law of photon index 2; interpreted as inverse-Compton and synchrotron emission, respectively.   

\item The Eddington ratio $L_{\rm x}/L_{\rm Edd}$ of OJ 287 is surprisingly low in X-rays. The low-state emission constrains any long-lived accretion-disk/corona contribution to $L_{\rm x}/L_{\rm Edd} \leq 5.6 \times 10^{-4} $
for $M_{\rm BH}=1.8\times10^{10}$ M$_{\odot}$, implying that the disk corona of OJ 287 is under-luminous in X-rays.  

\item We suggest that the 2016/17 X-ray outburst of OJ 287 is related to the 2016 optical outburst which was noted earlier and was interpreted as an after-flare predicted by the binary SMBH model.
Highest-resolution radio observations have the potential to locate the jet-emitting components within the nucleus
and follow their evolution in the future. 

\item Our most recent Swift observations revealed a deep low-state in September 2020, followed by a gradual recovery in flux. 

\end{itemize}

Given its brightness, proximity, and  candidate binary SMBH nature, OJ 287 is a remarkable multi-messenger source that can be well studied 
already, well 
before the actual direct detection of gravitational waves from supermassive black hole systems. 

\section*{Acknowledgements}
We would like to thank the Swift and  XMM-Newton teams for carrying out our observations and our anonymous referee for their useful comments and suggestions. 
JLG acknowledges financial support from the Spanish Ministerio de Econom\'{\i}a y Competitividad (grants AYA2016-80889-P, PID2019-108995GB-C21), the Consejer\'{\i}a de Econom\'{\i}a, Conocimiento, Empresas y Universidad of the Junta de Andaluc\'{\i}a (grant P18-FR-1769), the Consejo Superior de Investigaciones Cient\'{\i}ficas (grant 2019AEP112), and the State Agency for Research of the Spanish MCIU through the Center of Excellence Severo Ochoa award for the Instituto de Astrof\'{\i}sica de Andaluc\'{\i}a (SEV-2017-0709).
SGJ  acknowledges financial support from
NASA Fermi GI grants 80NSSC20K1565 and 80NSSC20K1566.
DH acknowledges funding from the Natural Sciences and Engineering Research Council of Canada (NSERC) and the Canada Research Chairs (CRC) program. 
We acknowledge the use of data we have obtained with the Neil Gehrels Swift mission. We also acknowledge the use of public data from the Swift data archive. 
This work made use of data supplied by the UK Swift Science Data Centre at the University of Leicester. 
This research is partly based on observations obtained with XMM-Newton, an ESA science mission with instruments and contributions directly funded by ESA Member States and NASA.
This research has made use of the
 XRT Data Analysis Software (XRTDAS) developed under the responsibility
of the ASI Space Science Data Center (SSDC), Italy. 

\section*{Data Availability Statement}
Reduced data are available upon reasonable request. Raw data can be retrieved from the Swift and XMM-Newton archives at \url{https://swift.gsfc.nasa.gov/archive/} and \url{https://www.cosmos.esa.int/web/xmm-newton}, respectively.  








\appendix

\bsp	
\label{lastpage}
\end{document}